\documentclass[final]{siamltex}

\usepackage{inputenc}
\usepackage{amsmath}
\usepackage{amssymb, amsfonts}
\usepackage{graphicx,epstopdf,bm}
\usepackage[colorlinks,bookmarks,urlcolor=blue,citecolor=blue,linkcolor=blue]{hyperref}
\usepackage{cite}
\usepackage[T1]{fontenc}
\usepackage{bbm}
\usepackage{mathtools}
\usepackage{subfig}

\usepackage{bbm}

\renewcommand{\vec}[1]{\boldsymbol{#1}}
\newcommand{\vecs}[1]{\boldsymbol{#1}}
\newcommand{\paren}[1]{\left(#1\right)}
\newcommand{\brac}[1]{\left[#1\right]}

\newcommand{\D}[2]{\frac{d#1}{d#2}}

\newcommand{\PD}[2]{\frac{\partial#1}{\partial#2}}
\newcommand{\PDD}[2]{\frac{\partial^{2}{#1}}{\partial{#2}^{2}}}

\newcommand{\lap}[1]{\Delta#1}

\newcommand{\abs}[1]{\left|#1\right|}
\newcommand{\norm}[1]{\Vert#1\Vert}

\DeclareMathOperator{\erfc}{erfc}

\newcommand{\comment}[1]{{\color{blue}{#1}}}

\newcommand{\kb}{k_{\text{B}}}

\newcommand{\veta}{\vecs{\eta}}

\newcommand{\vx}{\vec{x}}
\newcommand{\vy}{\vec{y}}

\newcommand{\vX}{\vec{X}}
\newcommand{\vV}{\vec{V}}
\newcommand{\vW}{\vec{W}}
\newcommand{\vv}{\vec{v}}

\newcommand{\vO}{\vec{0}}

\newcommand{\rb}{r_{\textrm{b}}}

\def\R{\mathbb{R}}

\newcommand{\plt}{\tilde{p}}

\newcommand{\pinf}{p_{\infty}}
\newcommand{\phibar}{\overline{\varphi}}
\newcommand{\pRob}{\overline{p}}
\newcommand{\dB}{\partial B}
\newcommand{\vxhat}{\hat{\vx}}
\newcommand{\vrhat}{\hat{v}_r}

\newcommand{\derphir}{\frac{\mbox{d}\varphi}{\mbox{d}r}}

\renewcommand{\comment}[1]{{#1}}

\title{Reactive Boundary Conditions as Limits of Interaction 
Potentials for Brownian \\ and Langevin Dynamics\thanks{The research 
leading to these results has received funding from the European 
Research Council under the \textit{European Community}'s Seventh 
Framework Programme ({\it FP7/2007-2013})/ ERC {\it grant agreement} 
n$^\circ$ 239870, and by National Science Foundation award 
DMS-1255408.}}  

\author{S. Jonathan Chapman\thanks{Mathematical Institute, 
University of Oxford, Radcliffe Observatory Quarter, 
Woodstock Road, Oxford, OX2 6GG, United Kingdom,
e-mail: chapman@maths.ox.ac.uk.}
\and Radek Erban\thanks{Mathematical Institute, 
University of Oxford, Radcliffe Observatory Quarter, 
Woodstock Road, Oxford, OX2 6GG, United Kingdom,
e-mail: erban@maths.ox.ac.uk. Radek Erban would 
like to thank the Royal Society for a University 
Research Fellowship and the Leverhulme Trust for 
a Philip Leverhulme Prize. This prize money was used 
to support research visits of Samuel Isaacson in Oxford.}
\and  
Samuel A. Isaacson\thanks{Department of Mathematics 
and Statistics, Boston University, 111 Cummington Mall, 
Boston, MA 02215, USA; e-mail: isaacson@math.bu.edu.}}

\numberwithin{equation}{section}

\begin{document}
\maketitle

\begin{abstract}
  \noindent A popular approach to modeling bimolecular reactions
  between diffusing molecules is through the use of reactive boundary
  conditions. One common model is the Smoluchowski partial adsorption
  condition, which uses a Robin boundary condition in the separation
  coordinate between two possible reactants. This boundary condition
  can be interpreted as an idealization of a reactive interaction
  potential model, in which a potential barrier must be surmounted
  before reactions can occur. In this work we show how the reactive
  boundary condition arises as the limit of an interaction potential
  encoding a steep barrier within a shrinking region in the particle
  separation, where molecules react instantly upon reaching the peak
  of the barrier. The limiting boundary condition is derived by the
  method of matched asymptotic expansions, and shown to depend
  critically on the relative rate of increase of the barrier height as
  the width of the potential is decreased. Limiting boundary
  conditions for the same interaction potential in both the overdamped
  Fokker-Planck equation (Brownian dynamics), and the Kramers equation
  (Langevin dynamics) are investigated. It is shown that different
  scalings are required in the two models to recover reactive boundary
  conditions that are consistent in the high friction limit (where the
  Kramers equation solution converges to the solution of the
  Fokker-Planck equation).
\end{abstract}




\pagestyle{myheadings}
\thispagestyle{plain}
\markboth{S. J. CHAPMAN, R. ERBAN, S. A. ISAACSON}{REACTIVE 
BOUNDARY CONDITIONS AS LIMITS OF INTERACTION POTENTIALS} 


\section{Introduction}

\noindent
Let $\vX(t) \in \Omega \subset \R^{d}$ and $\vV(t) \in \R^d$ denote 
the stochastic processes for position and velocity at time $t$ of 
a molecule moving in the $d$-dimensional domain $\Omega$ (where
$d \in {\mathbb N}$) according to the Langevin dynamics (LD):
\begin{equation}
\label{eq:langevinEq}
\begin{aligned}
\mbox{d} \vX(t) &= \vV(t) \, \mbox{d}t, \\
\mbox{d} \vV(t) &= 
-
\big( \beta \, \vV(t) + D \, \beta \, \nabla \varphi(\vX(t)) \big) 
\, \mbox{d}t 
+ \beta \, \sqrt{2 D} \, \mbox{d}\vW(t),
\end{aligned}
\end{equation}
where $\beta$ is the friction constant, $D$ is the diffusion constant,
$\varphi: \Omega \to {\mathbb R}$ is the potential (to be specified
later) and $\vW(t)$ denotes a $d$-dimensional Brownian motion. Here
$\beta$ is assumed to have units of `per time', $D$ units of `distance
squared per time', and $\varphi$ is assumed to be non-dimensional. In
physical units, the potential energy of a molecule at $\vx$ is then
$\kb T \varphi(\vx)$, \comment{where $\kb$ is the Boltzmann constant and
$T$ is the absolute temperature.} If $m$ denotes the mass of the molecule, 
the Einstein relation gives that $m D \beta = \kb T$, so that excepting
the noise term, \eqref{eq:langevinEq} follows from Newton's second law
of motion.

Passing to the overdamped limit $\beta \to \infty$ in
(\ref{eq:langevinEq}), we obtain the Brownian dynamics (BD) model for
$\vX(t)$ as follows:
\begin{equation}
\label{eq:BrownDyn}
\mbox{d} \vX(t) 
= D \, \nabla \varphi(\vX(t)) \, \mbox{d}t + \sqrt{2 D} \, \mbox{d}\vW(t).
\end{equation}
In the special case that $\varphi \equiv 0$ \comment{the molecule
  simply moves by Brownian motion,}
\begin{equation}
\label{BDdiff}
\mbox{d} \vX(t) = \sqrt{2 D} \, \mbox{d}\vW(t).
\end{equation}
Equation~(\ref{BDdiff}) is a popular description for the movement of
\comment{molecules within cells. It has been used to model spatial
  transport} in a number of computational packages for simulating
intracellular processes, including
Smoldyn~\cite{Andrews:2004:SSC,Andrews:2010:DSC,Robinson:2015:MRS},
GFRD~\cite{vanZon:2005:GFR,Takahashi:2010:STC} and
FPKMC~\cite{Opplestrup:2009:FKM,Mauro:2014fi}.  A common approach for
then coupling molecular interactions (diffusion-limited reactions)
to~\eqref{BDdiff} in these packages is to postulate that reactions can
occur \comment{by one of several possible mechanisms} if the
corresponding reactants are sufficiently
close~\cite{Erban:2009:SMR,Agbanusi:2014:CBR}.

The LD model~(\ref{eq:langevinEq}) with $\varphi \equiv 0$ provides a
more microscopic description of diffusion than the BD
model~(\ref{BDdiff}). It computes both position and velocity of a
molecule by assuming that the molecule is subject to a normally
distributed random force during each time increment. In particular, LD
can be considered as an intermediate description between detailed
molecular dynamics (MD) simulations and BD
simulators~\cite{Erban:2014:MDB}. Typical full-atom MD simulations use
time steps of the order of
$1 \; \mbox{fs} = 10^{-15} \; \mbox{s}$~\cite{Leimkuhler:2015:MDD},
while Smoldyn discretizes the equation (\ref{BDdiff}) with times steps
\comment{ranging from nanoseconds to milliseconds, depending on 
a particular application~\cite{Robinson:2015:MRS}. Stochastic
descriptions which compute both position and velocity of diffusing
particles, including LD, are applicable on intermediate time 
scales~\cite{Erban:2014:MDB,BDionref}}.

One advantage of Smoldyn or similar BD packages is that they can
simulate whole-cell dynamics. BD models based on
equation~(\ref{BDdiff}) have been applied to a number of biological
systems including signal transduction in {\it
  E. coli}~\cite{Lipkow:2005:SDP}, actin dynamics in
filopodia~\cite{Erban:2014:MSR}, the MAPK
pathway~\cite{Takahashi:2010:STC} and intracellular calcium
dynamics~\cite{Dobramysl:2015:PMM}.  In these applications, the
positions of all diffusing molecules are updated according
to~(\ref{BDdiff}) and the distances between each pair of possible
reactants (for bimolecular reactions) are calculated. Each reaction
then occurs (with a given probability) when the computed distance is
smaller than a \comment{specified reaction radius (as in Smoldyn), or
  alternatively occurs when the computed distance exactly equals a
  specified reaction-radius (as in GFRD and FPKMC)}.  To our
knowledge, there is no established spatio-temporal simulator of
intracellular processes based on the LD model.  In order to develop
one, one has to first investigate how bimolecular reactions might be
described in the LD context.

One possible way to implement bimolecular reactions in LD is to adopt
the same approach as in BD. That is, the positions and velocities of a
diffusing molecule would evolve according to the LD
model~(\ref{eq:langevinEq}) (with $\varphi \equiv 0$) and each
bimolecular reaction would occur (with a given probability) if the
distance between two reactants is smaller than the reaction
radius. However, as normally formulated this description of
bimolecular interactions would not make use of a molecule's velocity
(as is available in LD). That is, the LD bimolecular reaction model
would not provide any more physical detail than a BD model.

In this work we step back from the normal BD bimolecular reaction
model to the more microscopic reaction mechanism of a molecular
interaction potential. The general potential forms we consider
represent an irreversible bimolecular reaction as a molecule
surmounting a steep potential barrier in the separation coordinate
from a stationary target molecule, after which it enters an infinitely
deep well (the ``bound'' state). We show that the popular Smoluchowski
partial-adsorption BD reaction model~\cite{KeizerJPhysChem82} can be
derived in the simultaneous limit that the width of the barrier
approaches zero and the height of the barrier becomes infinite. Using
the same potential model, we then examine what limiting reactive
mechanism arises in the corresponding LD model, obtaining a specular
reflection boundary condition. We conclude by showing that in the high
friction limit, $\beta \to \infty$, the LD model with the specular
reflection bimolecular reaction model converges back to the
Smoluchowski partial-adsorption BD reaction model, consistent with the
kinetic boundary layer studies
of~\cite{Burschka:SelectAbsorpBC1981do,Burschka:FPKUnifField1982kr,Kneller:DiffControlRx1985}.
Our results demonstrate how an interaction potential model for a
bimolecular reaction can be parametrized in either LD or BD models to
be consistent with a BD model based on a partial-adsorption reaction
mechanism.

\section{Problem Setup}
We consider the movement of a molecule that can undergo an
irreversible bimolecular reaction with a second stationary molecule,
\comment{hereafter called the reactive target, which is modelled as 
a sphere of radius $\rb$. Let 
$B_{r}(\vO) \subset {\mathbb R}^d$ denote the $d$-dimensional 
ball of radius $r$ centered at the origin. Then the reactive target
is given as $\dB_{\rb}(\vO)$. We assume that the diffusing molecule 
moves within the $d$-dimensional domain $\Omega$ which satisfies
\begin{equation}
\Big( B_{L_1}(\vO) \setminus B_{\rb}(\vO) \Big) 
\; \subset \; 
\Omega,
\qquad \quad
\mbox{for} \quad \rb < L_1 < \infty.
\label{defomegagen1}
\end{equation} 
Equation~(\ref{defomegagen1}) defines $\Omega$ as the domain which is 
the exterior to the reactive target (sphere $\dB_{\rb}(\vO)$) 
and which includes all points which have a distance less than 
$L_1 - \rb > 0$ from the reactive target. A simple example of $\Omega$
satisfying~(\ref{defomegagen1}), for which we can find some 
explicit solutions in Section~\ref{secBDdis3}, is given as
\begin{equation}
\Omega =  {\mathbb R}^d \setminus B_{\rb}(\vO).
\label{defomegainf}
\end{equation} 
This is a standard ansatz for deriving formulae for the reaction
radius in BD descriptions, e.g. in the Smoluchowski or Doi
models~\cite{Smoluchowski:1917:VMT,Erban:2009:SMR,Agbanusi:2014:CBR}.
In most cases, we will further assume that $\Omega$ is bounded
with smooth boundary, i.e. equation~(\ref{defomegagen1}) is altered
by a bound from above
\begin{equation}
\Big( B_{L_1}(\vO) \setminus B_{\rb}(\vO) \Big) 
\; \subset \; 
\Omega
\; \subset \; 
B_{L_2}(\vO),
\qquad \quad
\mbox{for} \quad \rb < L_1 < L_2 < \infty.
\label{defomegagen}
\end{equation} 
A spherically symmetric example of domain $\Omega$ satisfying 
condition~(\ref{defomegagen}) is given as
\begin{equation}
\Omega =  B_{L}(\vO) \setminus B_{\rb}(\vO),
\qquad \quad
\mbox{for} \quad \rb < L < \infty.
\label{defomega}
\end{equation} 
If dimension $d=1$, then by the ``surface of the ball'' we mean the
origin, i.e. $\dB_{\rb}(0)=\{0\}$ and $\rb = 0$. Then
equation~(\ref{defomega}) reduces to the finite interval
$\Omega = \brac{0,L}$, which we use in our numerical examples.  

We assume that the molecule is adsorbed instantly upon reaching the
surface of the reactive target}, so that
\begin{equation}
\mbox{trajectory} \; \vX(t) \; \mbox{is terminated, if } 
\vX(t) \in \dB_{\rb}(\vO).  
\label{defadsorbinner}
\end{equation}
In order to reach the surface, the molecule has to overcome 
a potential barrier, which, denoting $r = \abs{\vx}$,  is given by 
\begin{equation}
\varphi(\vx) 
\equiv 
\varphi(r) 
= \left\{
\begin{matrix*}[l]
\displaystyle
\phibar  \, \psi \!\paren{ \frac{r-\rb}{\varepsilon}},
& \qquad \mbox{for} \; \rb \le r \le \rb + \varepsilon, \\ 
0, & \qquad \mbox{for} \; r > \rb + \varepsilon,
\end{matrix*}
\right.
\label{phidef} 
\end{equation}
where $\varepsilon > 0$ and $\phibar > 0$ are positive constants, and
$\psi: [0,1] \to [0,1]$ is a smooth function with a (unique) global
maximum at $\psi(0)=1$, and $\psi(1)=0$. These imply
\begin{equation*}
  \psi(z) < \psi(0), \quad 0 < z \leq 1,
\end{equation*}
with 
\begin{equation*}
\frac{\mbox{d} \psi}{\mbox{d} z}(0) < 0.
\end{equation*}
\comment{We assume that $\varepsilon$ in~(\ref{phidef}) is chosen
sufficiently small so that $\varepsilon \ll (L_1 - r_b)$ for $r_b$ and
$L_1$ given in assumption~(\ref{defomegagen1}).} 

Instead of studying the Langevin equation~\eqref{eq:langevinEq} we
shall work with the corresponding equation for the probability density
that $\vX(t) = \vx$ and $\vV(t) = \vv$, denoted $p(\vx,\vv,t)$, which
satisfies the Kramers equation (also called the phase-space
Fokker-Planck equation)
\begin{equation} 
\label{eq:kramersEq}
\PD{p}{t} + \vv \cdot \nabla_{\vx} p 
= 
\beta \, \nabla_{\vv} \cdot \brac{
\vv \, p + D \, p \, \nabla_{\vx} \varphi 
+ \beta \, D \, \nabla_{\vv} p }, 
\qquad \quad \mbox{for} \; \vx \in \Omega, \vv \in R^d,
\end{equation}
where $\nabla_{\vx}$ (resp.~$\nabla_{\vv}$) denotes the gradient
in $\vx$ (resp.~$\vv$) variable.
Considering the overdamped limit (\ref{eq:BrownDyn}), the
corresponding equation for the probability distribution
$p(\vx,t)$ is given as the Fokker-Planck equation
\begin{equation}
\PD{p}{t} 
= 
D \nabla_{\vx} \cdot \big[ 
\nabla_{\vx} p + p \, \nabla_{\vx} \varphi 
\big], \qquad \quad \mbox{for} \; \vx \in \Omega.  
\label{FPEq}
\end{equation}
In this paper, we show that equations (\ref{eq:kramersEq}) and
(\ref{FPEq}) with fully adsorbing boundary 
condition (\ref{defadsorbinner}) on $\dB_{\rb}(\vO)$ are 
in suitable limits equivalent
to a diffusion process with partially adsorbing (reactive, Robin)
boundary condition on $\dB_{\rb}(\vO)$, namely to the problem
\begin{equation} 
\label{diffplusrobin1}
\PD{p}{t} 
= 
D \, \lap_{\vx} p, \qquad \quad \mbox{for} \; \vx \in \Omega,
\end{equation}
with
\begin{equation} 
\label{diffplusrobin2}
D \, \nabla_{\vx} p (\vx) \cdot \frac{\vx}{\rb}
= 
K \, p(\vx),
\qquad \quad \mbox{for} \; \vx \in \dB_{\rb}(\vO),
\end{equation}
where $K$ is a suitable Robin boundary constant (reactivity of the
boundary) and $\vx/\rb$ is a unit normal vector to sphere
$\dB_{\rb}(\vO)$ at point $\vx$. Considering spherically symmetric
$\Omega$ given by~(\ref{defomegainf}) or (\ref{defomega}), and 
a spherically symmetric initial condition, we can rewrite the 
Fokker-Planck equation~(\ref{FPEq}) as
\begin{equation} 
\label{eq:diffWithGenPotential}
\PD{p}{t} 
= 
\frac{D}{r^{d-1}} 
\PD{}{r} 
\left(
r^{d-1}
\left[
\PD{p}{r}
+
p \, \derphir 
\right] 
\right),
\end{equation}
where $p(r,t)$ is the radial distribution function. Then the limiting Robin
boundary problem~(\ref{diffplusrobin1})--(\ref{diffplusrobin2}) is given as
\begin{equation} 
\label{diffplusrobin}
\PD{p}{t} 
= 
\frac{D}{r^{d-1}} 
\PD{}{r} 
\left(
r^{d-1}
\PD{p}{r} 
\right),
\qquad
\mbox{with}
\qquad
D \, \PD{p}{r}(\rb) 
= 
K \, p(\rb).
\end{equation}
The Robin boundary constant $K$ in equations (\ref{diffplusrobin2})
and (\ref{diffplusrobin}) is (together with $D$) an experimentally
determinable (macroscopic) parameter. In this way, we are able to
parametrize both BD and LD models using experimental data.

The rest of this paper is organized as follows. In
Section~\ref{secBD}, we investigate the BD
description~(\ref{eq:BrownDyn}). To get some insights into this
problem, we first consider the specific case of a linear interaction
potential and spherically symmetric domain~(\ref{defomegainf}) for $d=3$.
In this case, we can explicitly solve the corresponding Fokker-Planck
equation as shown in Section~\ref{secBDdis3}, and prove the
convergence of this solution, as $\varepsilon \to 0$ and
$\phibar \to \infty$, to the solution of a model involving a reactive
Robin boundary condition.  We then continue in Section~\ref{secBDgen}
with an asymptotic analysis, in the same dual limit, of the full BD
model~(\ref{eq:BrownDyn}) in general domain~(\ref{defomegagen}).  In
Section~\ref{secLD}, we investigate the LD model~(\ref{eq:langevinEq}) and
derive a boundary condition in the limit $\varepsilon \to 0$. This
boundary condition is then used in Section~\ref{secBDLD} to connect
the LD model to a BD model with Robin boundary condition in the dual
limits of high friction, $\beta \to \infty$, and large potential
barrier, $\phibar \to \infty$.  Numerical examples supporting our
analysis are provided in Sections~\ref{secnum1} and~\ref{secnum2}.  We
conclude with discussion in Section~\ref{secdiscussion}.

\section{Brownian Dynamics}
\label{secBD}

\noindent
In this section we consider the overdamped problem in which the
particle moves by BD, i.e. its position $\vX(t)$ evolves according to
equation~(\ref{eq:BrownDyn}) with the interaction potential
given by~(\ref{phidef}).

\subsection{Simple three-dimensional example with explicit solution}  
\label{secBDdis3}

Before investigating the general problem in $\R^d$, as a warm-up we
first consider a simplified special case to gain insight into how to
recover the Robin boundary condition (\ref{diffplusrobin}) from an
interaction potential. We consider the spherically symmetric
domain~(\ref{defomegainf}) for $d=3$. The steady-state
spherically-symmetric Fokker-Planck equation
(\ref{eq:diffWithGenPotential}) is then given by
\begin{equation}
0 = \frac{D}{r^2} 
\PD{}{r} \paren {r^2 \brac{ \PD{p}{r} + p \, \derphir }}, 
    \qquad \quad \mbox{for} \quad \rb < r < \infty,
\label{ststFPd3}
\end{equation}
where the fully adsorbing boundary condition (\ref{defadsorbinner})
implies a Dirichlet boundary condition at $r=\rb$, i.e. $p(\rb) = 0.$
We consider constant concentration $\pinf > 0$ far away from the
reactive surface, i.e.\footnote{Of course, $p$ as defined here is
  not a probability distribution, since it does not integrate to unity
  on the infinite domain $r>r_{\rm b}$. The problem as stated arises 
from a limiting process in which $p$ is re-scaled appropriately as the
domain size tends to infinity.} 
\begin{equation}
\lim_{r \to \infty} p(r) = \pinf.
\label{pinfdef}
\end{equation}
We also assume that the potential $\varphi(r)$ is given
by~(\ref{phidef}) where $\psi: [0,1] \to [0,1]$ is the linear function
\begin{equation}
  \label{eq:linPotential}
  \psi(z) = 1-z.
\end{equation}
Then
\begin{equation*}
  \derphir(r) \equiv
\begin{cases}
\displaystyle
 - \frac{\phibar}{\varepsilon}, & \rb < r < \rb + \varepsilon, \\
0, & \rb + \varepsilon < r. 
\end{cases}
\end{equation*}
\comment{Substituting into~(\ref{ststFPd3}), and making use of the boundary
conditions at $r=\rb$ and $r=\infty$, we can solve piecewise to obtain
a solution on $(\rb,\rb+\varepsilon)$ and a solution on
$(\rb + \varepsilon, \infty)$. On each interval the solution is
defined up to an unknown constant. By enforcing continuity of the flux
at $r = \rb$ we can eliminate one constant to obtain}
\begin{equation} 
\label{eq:potentSolut}
p(r) = 
\left\{
\begin{matrix*}[l]
\displaystyle
A \int_{\rb}^{r} \frac{1}{s^2}
\exp\left[\displaystyle \frac{\phibar (r-s)}{\varepsilon} \right]\, \mbox{d} s,
& \qquad \mbox{for} \; \rb \le r \le \rb + \varepsilon, \\ 
\displaystyle
\pinf - \frac{A}{r}, & \qquad \mbox{for} \; r > \rb + \varepsilon,
\end{matrix*}
\right.
\end{equation}
where the last constant, $A$, is determined by the continuity of
$p(r,t)$ at $r = \rb + \varepsilon$. This gives
\begin{equation*}
 A 
 = 
 \frac{(\rb + \varepsilon) \, \pinf}{1 + (\rb + \varepsilon)
 \beta(\varepsilon,\phibar)},
\end{equation*}
with 
\begin{equation}
\beta(\varepsilon,\phibar) 
= \varepsilon \int_{0}^{1} 
\frac{\exp[ \, \phibar(1-s) \, ]}{(\rb + \varepsilon s)^2} \, \mbox{d}s.
\label{defbeta}
\end{equation}
To recover the Robin condition (\ref{diffplusrobin}) as
$\varepsilon \to 0$ and $\phibar \to \infty$, we need
\begin{equation*}
\lim_{\substack{\varepsilon \to 0 \\ \phibar \to \infty}}
\; \lim_{r \to (\rb + \varepsilon)^+}
\brac{D \PD{p}{r}(r) 
- K p(r)} = 0,
\end{equation*}
which is equivalent to
\begin{equation*}
\lim_{\substack{\varepsilon \to 0 \\ \phibar \to \infty}} 
\frac{D + K (\rb +\varepsilon)}
{(\rb + \varepsilon) 
\big(1 + (\rb + \varepsilon) \beta(\varepsilon,\phibar) \big)} = K,
\end{equation*}
or simplifying,
\begin{equation}
\lim_{\substack{\varepsilon \to 0 \\ \phibar \to \infty}}
\beta(\varepsilon,\phibar)
= \frac{D}{K \rb^2}.
\label{desiredlimit}
\end{equation}
We note that
\begin{align} \label{eq:betaBound}
0 \leq \beta(\varepsilon,\phibar)
\leq \frac{\varepsilon}{\rb^2} 
\int_{0}^{1} \exp[ \, \phibar(1-s) \, ]\, \mbox{d}s 
= \frac{\varepsilon \, (\exp[ \, \phibar \, ]-1)}{\rb^2 \, \phibar}. 
\end{align}
Therefore, $\beta(\varepsilon,\phibar)$ will have a finite limit 
if $\varepsilon \exp[ \, \phibar \, ] / \phibar$ has a finite limit. 
This motivates the choice
\begin{equation} \label{epsphi}
\varepsilon = \frac{D \, \phibar \exp[ \, - \phibar \, ]}{K} .
\end{equation}
Using~(\ref{defbeta}), we have
\begin{eqnarray} 
\nonumber
\abs{\beta(\varepsilon,\phibar) - \frac{\varepsilon}{\rb^2} 
\int_{0}^{1} \exp[ \, \phibar(1-s) \, ]\, \mbox{d}s}
&=&
\abs{\varepsilon 
\int_{0}^{1} 
\exp[ \, \phibar(1-s) \, ]
\left(
\frac{1}{(\rb + \varepsilon s)^2}
- 
\frac{1}{\rb^2} 
\right) \, \mbox{d}s }
\\
&\le&
\frac{D \, \varepsilon \, (2 \rb + \varepsilon)}{K \, \rb^4},
\label{eq:betaErr}
\end{eqnarray}
which converges to zero as $\varepsilon \to 0$. 
We therefore conclude that
\begin{eqnarray*}
\lim_{\substack{\varepsilon \to 0, \;\phibar \to \infty, \\
\mbox{\scriptsize where} \; \varepsilon \;\mbox{\scriptsize and}\; 
\phibar \;\mbox{\scriptsize are} \\ 
\mbox{\scriptsize related by~(\ref{epsphi})}  }}
\!\!\!\beta(\varepsilon,\phibar)
&=& 
\lim_{\substack{\varepsilon \to 0, \;\phibar \to \infty, \\
\mbox{\scriptsize where} \; \varepsilon \;\mbox{\scriptsize and}\; 
\phibar \;\mbox{\scriptsize are} \\ 
\mbox{\scriptsize related by~(\ref{epsphi})}  }}
\frac{\varepsilon}{\rb^2} \int_{0}^{1} \exp[ \, \phibar(1-s) \, ] \, \mbox{d}s 
\\
&=&
\lim_{\phibar \to \infty} 
\frac{D \, \phibar \exp[ \, - \phibar \, ]}{K \, \rb^2}
\int_{0}^{1} \exp[ \, \phibar(1-s) \, ] \, \mbox{d}s 
=
\frac{D}{K \rb^2}, 
\end{eqnarray*}
so that we get (\ref{desiredlimit}). In particular, we
recover the Robin boundary condition with Robin constant $K$. 

The steady-state solution to the limiting Robin boundary condition
problem~(\ref{diffplusrobin}) with~(\ref{pinfdef}) is given by
\begin{equation*}
\pRob(r) = \pinf \brac{ 1 - \frac{1}{1 + \dfrac{D}{K \rb}} \frac{\rb}{r}},
\qquad \quad \mbox{for} \quad \rb < r < \infty.
\end{equation*}
We now examine the error between this limit and 
the solution~\eqref{eq:potentSolut} 
for two cases: $\rb \le r \le \rb + \varepsilon$ 
and $r > \rb + \varepsilon$. For the latter we have
\begin{equation*}
  \abs{p(r) - \pRob(r)} 
  =  \frac{\pinf \, \abs{ \,
  \varepsilon \paren{1 + \frac{D}{K \rb} - \rb \, \beta(\varepsilon,\phibar)}
 + 
 \rb^2 \paren{ \frac{D}{K \rb^2 } - \beta(\varepsilon,\phibar) }}}
 {r \Big( 1 + (\rb + \varepsilon)
 \beta(\varepsilon,\phibar) \Big) \left( 1 + \frac{D}{K \rb} \right)}.
\end{equation*}
We can estimate that the denominator is greater than
$r$. Consequently,
\begin{equation*}
  \abs{p(r) - \pRob(r)} 
  <  \frac{\pinf \, \varepsilon}{r} \, 
  \paren{1 + \frac{D}{K \rb} + \rb \, \beta(\varepsilon,\phibar)} 
 + 
 \frac{\pinf \, \rb^2}{r} \abs{ \frac{D}{K \rb^2 } - \beta(\varepsilon,\phibar) }.
\end{equation*} Using (\ref{eq:betaBound}) and the definition 
of $\varepsilon$ (equation~\eqref{epsphi}), we have
$\beta(\varepsilon,\phibar) \sim O(1)$ as $\varepsilon \to 0$.
Using~\eqref{eq:betaErr}, we conclude that the second term on 
  the right hand side is $O(\varepsilon/r)$ as $\varepsilon \to 0$.
  Thus, we conclude
  \begin{equation} \label{eq:ppRobErr}
    \abs{p(r) - \pRob(r)} = O\paren{\frac{\varepsilon}{r}} = O(\varepsilon),
    \qquad \quad \mbox{for} \quad r > \rb + \varepsilon.
  \end{equation}  
  On the other hand, we have
\begin{align*}
  \sup_{r \in (\rb,\rb + \varepsilon)} \abs{p(r) - \pRob(r)} 
  \geq \abs{p(\rb) - \pRob(\rb)} 
  = 
  \pRob(\rb) 
  = 
  \frac{\pinf D}{K \rb + D} > 0.
\end{align*}
The maximum error between the two models is therefore $O(1)$ as
$\varepsilon \to 0$ (where $\phibar$ satisfies~(\ref{epsphi})),
illustrating the non-uniformity of the error within the region where 
the potential interaction is non-zero.

  Note, while the maximum norm (i.e. $L_{\infty}$ norm) of the
  difference between solutions does not converge on
  $(\rb,\rb + \varepsilon)$, both $p(r)$ and $\pRob(r)$ are uniformly
  bounded on $(\rb,\rb+\varepsilon)$ since
  \begin{align*}
    \sup_{r \in (\rb,\rb+\varepsilon)} p(r) &\leq A \, \beta(\varepsilon,\phibar) 
    \leq \frac{2 \, D \, \pinf}{K \, \rb},\\
    \sup_{r \in (\rb,\rb+\varepsilon)} \pRob(r) &\leq \pinf.
  \end{align*}
  As such, using that the maximum norm error is $O(\varepsilon/r)$ on
  $\paren{\rb+\varepsilon,\infty}$ by~\eqref{eq:ppRobErr}, we find
  that the $q$-norm converges for any $3 < q < \infty$,
  \begin{equation*}
    \norm{p(r) - \pRob(r)}_q
    = \paren{\int_{\rb}^{\infty} \abs{p(r)-\pRob(r)}^q r^2 \, dr}^{\frac{1}{q}} 
    = O\paren{\varepsilon^{\frac{1}{q}}}.
  \end{equation*}
  Here, the lower bound $q > 3$ is an artifact of our working on an
  unbounded domain, which requires integrability of
  $\abs{p(r) - \pRob(r)}^q r^2$ on $(\rb,\infty)$. If our domain was
  bounded, i.e. given by equation~(\ref{defomega}), and we used the
  Dirichlet boundary condition that $p(L) = \pinf$, we would expect a
  similar estimate to hold for all $1 \leq q < \infty$. 

  \comment{In summary, we have found that in the special case of a
    linear potential barrier, by taking the height of the barrier
    $\phibar \to \infty$ and then choosing the width of the barrier,
    $\varepsilon$, to decrease exponentially in the height
    (i.e. according to equation~(\ref{epsphi})), we can
    recover the solution to the diffusion equation with Robin boundary
    condition. We may therefore interpret the Robin boundary condition
    model for bimolecular reactions between two molecules as
    approximating an underlying interaction potential, in which the
    two molecules must surmount a steep potential barrier before
    entering a bound state represented by a deep well.}

\vskip 1.4mm

\noindent {\bf Remark.}
{\rm In~{\rm \cite{SzaboShoup1982}} it was shown how 
the Robin constant can be chosen to give the same 
diffusion-limited reaction rate in two steady-state 
spherically-symmetric models both including the same
\emph{fixed} interaction potential, with one using 
a zero Dirichlet boundary condition at $\rb$ 
(as in~\eqref{ststFPd3}), and the other using a 
Robin condition at $\rb + \varepsilon$ for any
fixed $\varepsilon > 0$.  The argument 
in~{\rm \cite{SzaboShoup1982}} can be modified to match 
diffusion-limited reaction rates in the
two models considered in this section (i.e. \eqref{ststFPd3} with
$p(\rb)=0$, and the steady-state diffusion equation with Robin
boundary condition satisfied by $\bar{p}(r)$).  It is satisfying
to note that choosing $K$ to match the diffusion limited rates in
this manner also gives the relationship~\eqref{epsphi} between
$K$, $\varepsilon$ and $\phibar$. We therefore conclude that
choosing $K$ to match the diffusion-limited reaction rates of the
two models of this section is sufficient to give convergence of
$p(r)$ to $\bar{p}(r)$ as $\varepsilon \to 0$ and
$\phibar \to \infty$.}
 
\subsection{Asymptotics for a general BD model with a general interaction
potential}
\label{secBDgen}

\noindent
To analyze equation~(\ref{FPEq}) for arbitrary 
$d \in {\mathbb N}$ with general potential $\varphi(r)$ in the 
form (\ref{phidef}), we follow the approach of Pego~\cite{pego:1989wf} 
and Li~\cite{li:2009vs}. We use the method of matched asymptotic
expansions in a general bounded domain $\Omega$ which satisfies
(\ref{defomegagen}) and has a smooth boundary.
First, we consider an inner solution in 
a boundary layer near $\dB_{\rb}(\vO)$. Let
\begin{equation}
z = \frac{r-\rb}{\varepsilon},
\label{varzdef}
\end{equation}
denote the stretched distance of a point $\vx$ from the reactive
boundary. We define an inner solution $\plt(z,\vx,t)$. The second
argument should actually be $\vxhat = \vx / \abs{\vx}$, but we follow
the method of~\cite{pego:1989wf} and instead assume that
\begin{equation*}
  \plt(z,\lambda \vx,t) = \plt(z,\vx,t), 
\end{equation*}
for all $\lambda > 0$ (i.e. the length of $\vx$
is accounted for by $z$). We note the identities that
\begin{equation} \label{eq:divGradxIdents}
  \begin{aligned}
    \nabla_{\vx} \cdot \vxhat &= \frac{d-1}{r} 
    = \frac{d-1}{\rb + \varepsilon z}, \\
    \nabla_{\vx} z &= \frac{\vxhat}{\varepsilon}.
  \end{aligned}
\end{equation}
In the new coordinate system 
\begin{align*}
  \nabla_{\vx} p &
  \to \nabla_{\vx} \plt + \frac{1}{\varepsilon} \PD{\plt}{z} \vxhat, \\
  \lap_{\vx} p &
  \to \lap_{\vx} \plt + \frac{1}{\varepsilon}
  \frac{d-1}{\rb + \varepsilon z} \PD{\plt}{z} 
  + \frac{1}{\varepsilon^2} \PDD{\plt}{z},
\end{align*}
where we have used~\eqref{eq:divGradxIdents} and that
\begin{equation*}
  \vxhat \cdot \nabla_{\vy} \PD{\plt}{z}(z,\vy,t)\Big|_{\vy=\vx} = 0,
\end{equation*}
as $\plt$ is assumed constant when the second argument is varied along
the $\vxhat$ direction. From~\eqref{FPEq} we find the inner problem
(for $0< z < 1$)
\begin{equation*}
\PD{\plt}{t} 
= 
D \left[
\lap_{\vx} \plt + \frac{1}{\varepsilon}
  \frac{d-1}{\rb + \varepsilon z} \PD{\plt}{z} 
  + 
  \frac{1}{\varepsilon^2} \PDD{\plt}{z}
  + 
  \frac{\phibar}{\varepsilon^2} \PD{\plt}{z} 
  \frac{\mbox{d} \psi}{\mbox{d} z}
  +
  \frac{\phibar}{\varepsilon}
  \frac{d-1}{\rb + \varepsilon z} 
  \, p \, \frac{\mbox{d} \psi}{\mbox{d} z} 
  + 
  \frac{\phibar}{\varepsilon^2} 
  \, p \, \frac{\mbox{d}^2 \psi}{\mbox{d} z^2}
\right]. 
\end{equation*}
We now expand the inner and outer solutions in $\varepsilon$ and denote the
leading-order terms by $\plt_0$ and $p_0$ respectively. The 
leading-order behavior of the inner solution then satisfies
\begin{equation*}
  \PDD{\plt_0}{z}
  + 
  \phibar \, \PD{\plt_0}{z} 
  \frac{\mbox{d} \psi}{\mbox{d} z}
  + 
  \phibar \, 
  \, \plt_0 \, \frac{\mbox{d}^2 \psi}{\mbox{d} z^2}
 = 0. 
\end{equation*}
 Note that we will find below that in the distinguished limit
 $\phibar$ depends on $\varepsilon$ in a logarithmic manner (see
 equation \eqref{epsscalinggenBD}) so that retaining $\phibar$ here is
 consistent with neglecting terms of $O(\varepsilon)$.
Solving this equation and using the Dirichlet boundary condition at
$z=0$, we find
\begin{equation*}
\plt_0(z,\vx,t) 
= 
A(\vx,t) 
\int_0^z 
\exp \! \big[ \,
\phibar \, (\psi(z')-\psi(z))
\big] \, \mbox{d}z',
\end{equation*}
where $A(\vx,t)$ is an unknown constant, also satisfying the dilation
property 
\begin{equation*}
A(\lambda \vx, t) 
= A(\vx,t), \qquad \quad \mbox{for} \quad \lambda > 0. 
\end{equation*}
The leading order term of the outer solution, $p_0$, satisfies the
diffusion equation~(\ref{diffplusrobin1}) away from the reactive
boundary.  Our matching conditions are 
\begin{align*}
\lim_{\vx \to \dB_{\rb}(\vO)} p_0(\vx,t) 
&
= \lim_{z \to 1} \plt_0(z,\vx,t) 
= A(\vx,t) 
\int_0^1 
\exp \! \big[ \,
\phibar \, \psi(z')
\big] \, \mbox{d}z', \\
\lim_{\vx \to \dB_{\rb}(\vO)} \nabla p_0(\vx,t) \cdot \vxhat &= 
\lim_{\vx \to \dB_{\rb}(\vO)} \PD{p_0}{r} 
= 
\lim_{z \to 1} \frac{1}{\varepsilon} 
\brac{
\PD{\plt_0}{z}  + \phibar \, \plt_0 
\, \frac{\mbox{d} \psi}{\mbox{d} z} 
}
= \frac{A(\vx,t)}{\varepsilon}.
\end{align*}
Combining these two equations, for $\vx \in \dB_{\rb}(\vO)$ we find 
\begin{equation*}
D \, \nabla_{\vx} p (\vx) \cdot \frac{\vx}{\rb}
= 
D \PD{p_0}{r}(\vx,t) 
= 
\frac{D \,p_0(\vx,t)}{\varepsilon 
\int_0^1 
\exp \! \big[ \,
\phibar \, \psi(z')
\big] \, \mbox{d}z'
}
\sim
\frac{D \, \phibar}{\varepsilon \exp(\phibar)}
\abs{\frac{\mbox{d} \psi}{\mbox{d} z}(0)}
\,
p_0(\vx,t),
\end{equation*}
where the last approximation is obtained using Laplace's method
in the limit $\phibar \to \infty.$ Thus we recover the desired 
Robin boundary condition (\ref{diffplusrobin2}) if
$\phibar \to \infty$ and $\varepsilon \to 0$ such that
\begin{equation}
\varepsilon = \frac{D \, \phibar}{K \exp(\phibar)}
\abs{\frac{\mbox{d} \psi}{\mbox{d} z}(0)}.
\label{epsscalinggenBD}
\end{equation}

\comment{In summary, we have used the method of matched asymptotic
  expansions to examine scaling limits for a general set of potential
  interactions between two particles.  The potentials were assumed to
  be short-range, forming a high barrier as the separation between the
  molecules approaches a fixed ``reaction-radius'', and then a deep
  well once this barrier is surmounted. In the limit that the height
  of the barrier, $\phibar \to \infty$, and the width of the barrier,
  $\varepsilon$, decreases to zero exponentially in the height, we
  recover the solution to the diffusion equation with Robin boundary
  condition. We may therefore interpret bimolecular reactions between
  two molecules modeled with a Robin boundary condition as an
  approximation to one of many possible underlying potential
  interactions. These interactions are characterized by the two
  molecules needing to surmount a steep potential barrier before
  entering a bound state represented by a deep well.}

\vskip 1.4mm

\noindent{\bf Remark.}
If $\varepsilon$ is chosen to approach zero slower 
than~(\ref{epsscalinggenBD}), then we recover
a zero Neumann boundary condition at the reactive surface,
\begin{equation}
  \PD{p_0}{r}(\vx,t) = 0, 
  \qquad \quad \mbox{for} \; \vx \in \dB_{\rb}(\vO).
\label{neumannboundary}  
\end{equation}
Likewise, if $\varepsilon$ is chosen to approach zero 
faster than~(\ref{epsscalinggenBD}), then  we
recover the zero Dirichlet boundary condition that
\begin{equation*}
  p(\vx,t) = 0, \qquad \quad \mbox{for} \; \vx \in \dB_{\rb}(\vO). 
\end{equation*}

\section{Langevin Dynamics}
\label{secLD}

\noindent
We now consider the LD model~(\ref{eq:langevinEq}) in a bounded
$d$-dimensional domain satisfying~(\ref{defomegagen}).  The reactive
target is again taken to be the surface of the $d$-dimensional sphere
of radius $\rb$ about the origin.  It is again assumed that the
molecule is adsorbed instantly upon reaching the surface of the
sphere, i.e. we consider the boundary condition (\ref{defadsorbinner})
together with spherically symmetric interaction potential $\varphi$
given by (\ref{phidef}).
We leave unspecified any boundary condition on
$\partial \Omega \setminus \dB_{\rb}(\vO)$, as it is not needed in the
following analysis.  

Instead of studying the Langevin equation~\eqref{eq:langevinEq} we
work with the corresponding Kramers equation~(\ref{eq:kramersEq}).
Since $\vx/\rb$ is the normal to $\dB_{\rb}(0)$ at $\vx$, the
adsorbing boundary condition (\ref{defadsorbinner}) means that the
Kramers equation~\eqref{eq:kramersEq} is coupled with the Dirichlet
boundary condition
\begin{equation}
p(\vx,\vv,t) = 0
\qquad \quad \mbox{for} \; 
\vx \in \dB_{\rb}(0)
\; \mbox{and} \; 
\vv\cdot \vx > 0,
\label{Kramersboundary}
\end{equation}
and the unspecified boundary condition on
$\partial \Omega \setminus \partial B_{\rb}(\vO)$.  We are interested
in various limits of~(\ref{eq:kramersEq}) as $\varepsilon \to 0$,
$\phibar \to \infty$ and $\beta \to \infty$ \comment{in which the
interaction potential can be approximated by an appropriate reactive
boundary condition. In the BD models of 
Section~\ref{secBD} our goal was to derive the widely-used Robin 
boundary condition. To our knowledge, in the LD model we now 
investigate there is no standard reactive boundary condition for 
bimolecular reactions. Therefore, we wish to see what, if any, 
reactive boundary condition arises when considering similar 
limits of $\varepsilon$ and $\phibar$.}  

As in Section~\ref{secBDgen}, to study these limits we will match 
an inner solution within an $O(\varepsilon)$ boundary layer about
$\dB_{\rb}(\vO)$ to an outer solution when far from the reactive
boundary. Using the same notation as in Section~\ref{secBDgen}, 
we denote by $z$ the stretched distance from the boundary, given
by~(\ref{varzdef}). We also introduce a re-scaled radial velocity,
\begin{equation*}
  \nu = \frac{v_r}{\vrhat} = \frac{\vv \cdot \vxhat}{\vrhat},
\end{equation*}
where $\vxhat = \vx / \abs{\vx}$ is a unit vector in the $\vx$
direction, $v_r = \vv \cdot \vxhat$ is the radial velocity into
$\Omega$ from $\partial B_{\rb}(\vO)$, and $\vrhat$ is a scaling
constant in the radial velocity that will be specified later. In the
inner region we denote the density by $\plt(z,\vx,\nu,\vv,t)$, where
we assume that $\plt$ is constant whenever $\vx$ is varied in the
radial direction, and when the component of the velocity, $\vv$, in
the radial direction is varied. That is
\begin{align*}
  \plt(z,\vx,\nu,\vv,t) &= 
  \plt(z, \lambda \vx, \nu, \vv, t), 
  \qquad \qquad \qquad \qquad \mbox{for} \; \lambda > 0, \\
  \plt(z,\vx,\nu,\vv,t) &= 
  \plt(z, \vx, \nu, \vv + \alpha \vxhat, t), 
  \qquad \qquad \qquad \,\! \mbox{for} \; \alpha \in {\mathbb R}. 
\end{align*}
In addition to the identities~\eqref{eq:divGradxIdents}, we note that
\begin{equation}
\nabla_{\vx} \nu 
= 
\frac{\vv}{\vrhat (\rb + \varepsilon z)}
-
\frac{\nu \, \vxhat}{\rb + \varepsilon z}.
\label{eqgradvident}
\end{equation}
Using~\eqref{eq:divGradxIdents} and (\ref{eqgradvident})  we find 
the derivative operators in~(\ref{eq:kramersEq}) transform 
in the new coordinates to
\begin{align*}
 \nabla_{\vx} p &\to  
 \nabla_{\vx} \plt + \frac{1}{\varepsilon} \PD{\plt}{z} \vxhat 
 + 
\left(
\frac{\vv}{\vrhat (\rb + \varepsilon z)}
-
\frac{\nu \, \vxhat}{\rb + \varepsilon z}
\right)
 \PD{\plt}{\nu}, \\
\nabla_{\vv} p &\to 
\nabla_{\vv} \plt + \frac{1}{\vrhat} \PD{\plt}{\nu} \vxhat, \\
\lap_{\vv} p &\to 
\lap_{\vv} \plt + \frac{1}{\vrhat^2} \PDD{\plt}{\nu},
\end{align*}
where we have used that
\begin{equation*}
\vxhat \cdot \nabla_{\vv'} \PD{\plt}{\nu}(z,\vx,\nu,\vv',t)\Big|_{\vv'=\vv} 
= 0. 
\end{equation*}
Using~(\ref{phidef}), equation~(\ref{eq:kramersEq}) can be 
transformed to
\begin{multline*}
\PD{\plt}{t} 
+ 
\vv \cdot \nabla_{\vx} \plt 
+ 
\frac{\vrhat \, \nu}{\varepsilon} \PD{\plt}{z} 
 + 
\left(
\frac{\vv \cdot \vv}{\vrhat (\rb + \varepsilon z)}
-
\frac{\vrhat \, \nu^2}{\rb + \varepsilon z}
\right)
 \PD{\plt}{\nu}
= \\
\beta \Bigg( 
d \plt + \vv \cdot \nabla_{\vv} \plt + \nu \PD{\plt}{\nu}  
+ 
\frac{\phibar \, D}{\vrhat \, \varepsilon} \, 
\frac{\mbox{d} \psi}{\mbox{d} z}
\, \PD{\plt}{\nu} 
+ 
\beta D \lap_{\vv} \plt + \frac{\beta D}{\vrhat^2} \PDD{\plt}{\nu} 
\Bigg). 
\end{multline*}
We consider an asymptotic expansion of the inner solution, $\plt$ (in
the boundary layer about $\dB_{\rb}(\vO)$) as
$\varepsilon \to 0$ with all other parameters held fixed,
\begin{equation*}
\plt \sim  \plt^{(0)} + \plt^{(1)}\varepsilon + \plt^{(2)} \varepsilon^2 + \dots.
\end{equation*}
Similarly we also consider an expansion of the outer solution, valid
far from $\dB_{\rb}(\vO)$, for which we will abuse notation and also
denote it by $p$,
\begin{equation*}
  p \sim p^{(0)} + p^{(1)} \varepsilon + p^{(2)} \varepsilon^2 + \dots. 
\end{equation*}
At leading order $O(\varepsilon^{-1})$ we find that 
the inner solution satisfies
\begin{equation}
\vrhat \, \nu \, \PD{\plt^{(0)}}{z} 
-
\frac{\phibar \, \beta \, D}{\vrhat} \, 
\frac{\mbox{d} \psi}{\mbox{d} z} \,
\PD{\plt^{(0)}}{\nu} 
= 
0.
\label{eq:zeroOrderKramersExp1}
\end{equation}
To simplify this equation, we let 
\begin{equation}
\vrhat = \sqrt{\phibar \, \beta \, D}.
\label{vrhatchoice}
\end{equation}
This choice emphasizes large velocities, for which we expect the particle 
to be most likely to escape over the (reactive) potential barrier.
Substituting~(\ref{vrhatchoice}) into~(\ref{eq:zeroOrderKramersExp1}),
we obtain
\begin{equation} 
\label{eq:zeroOrderKramersExp}
\nu \PD{\plt^{(0)}}{z} 
 - 
 \frac{\mbox{d} \psi}{\mbox{d} z} \,
\PD{\plt^{(0)}}{\nu} = 0.
\end{equation}
The boundary condition~(\ref{Kramersboundary}) implies
\begin{equation} \label{eq:zeroOrderBC}
\plt^{(0)}(0,\vx,\nu,\vv,t) = 0, 
\qquad \mbox{for} \;  \nu > 0.
\end{equation}%
\begin{figure}[t]
\relax
\vskip 2mm
\centerline{
\hskip 2mm
\includegraphics[width=0.48\columnwidth]{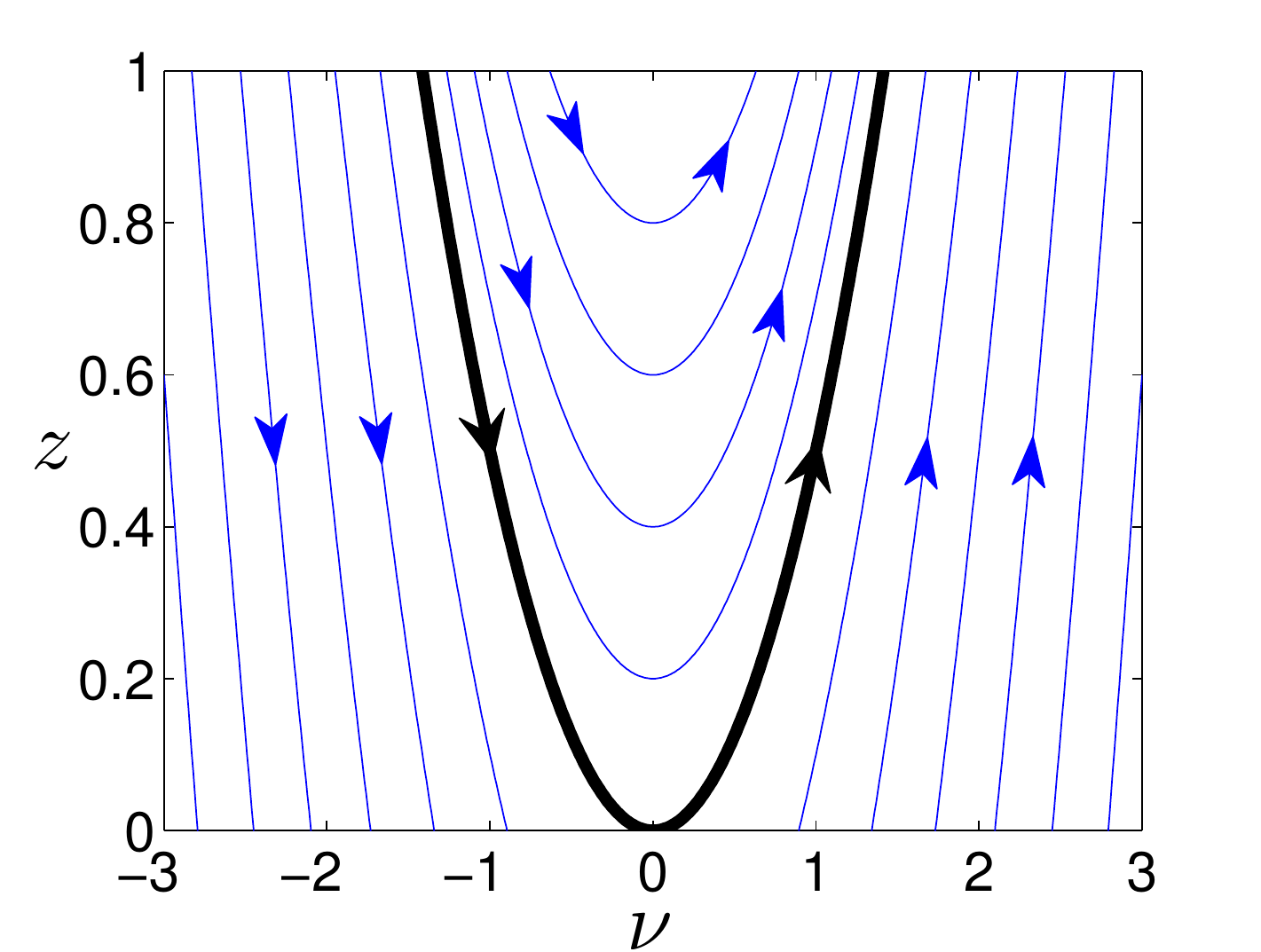}
\hskip 4mm
\includegraphics[width=0.48\columnwidth]{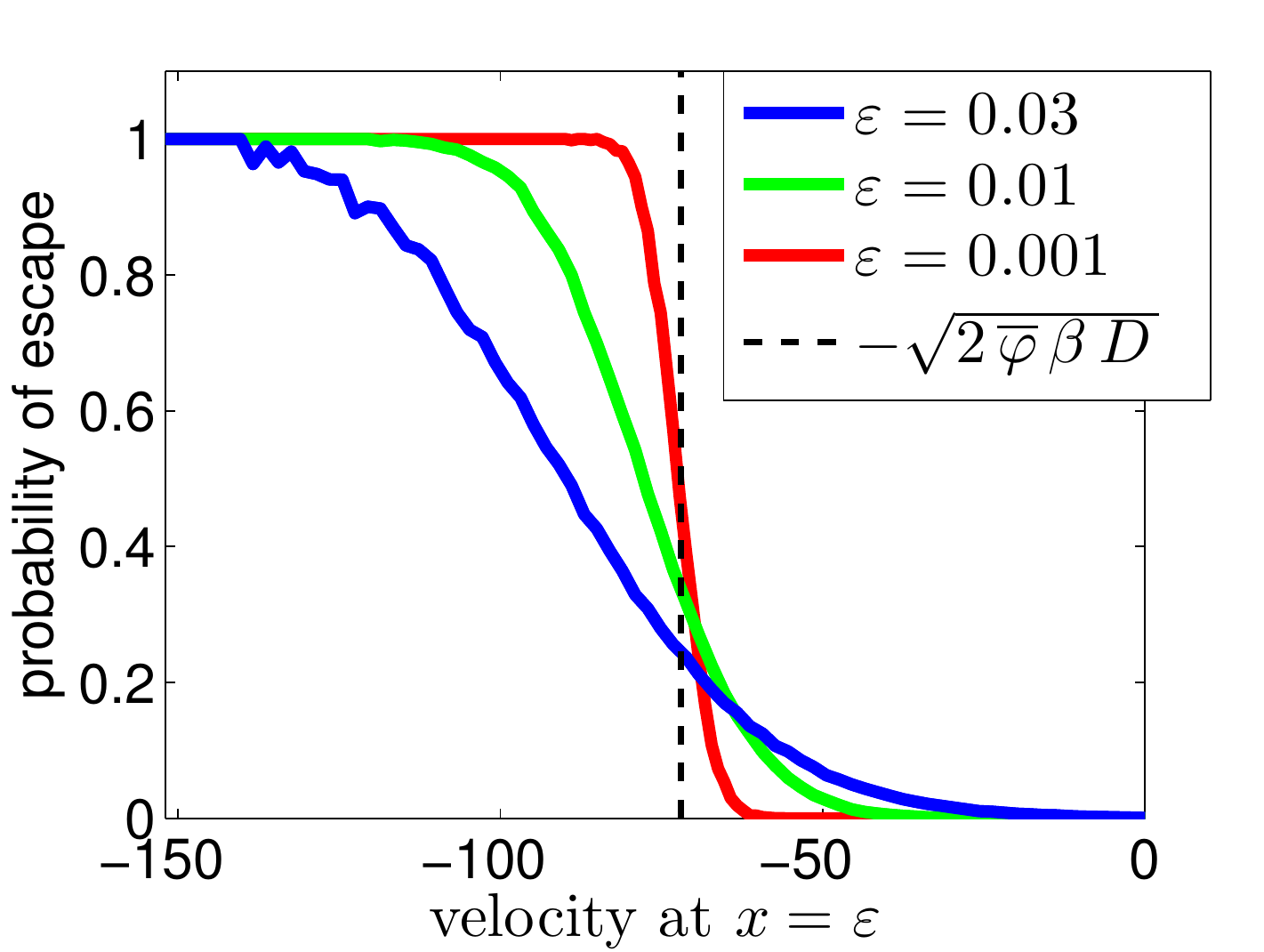}
}
\vskip -4.9cm
\leftline{\hskip 1mm (a) \hskip 6.1cm (b)}
\vskip 4.3cm
\caption{{\rm (a)} {\it Characteristic curves of}~{\rm
      (\ref{eq:zeroOrderKramersExp})} {\it for the linear
      potential}~{\rm (\ref{eq:linPotential})} {\it are shown as blue
      lines. The curve $z = \nu^2/2$ is drawn as a thicker black line,
      dividing the upper half plane into a region where the moving
      particle will escape over the barrier ($\nu < 0$, with $z <
      \nu^2/2$), a region where all trajectories are reflected ($z >
      \nu^2/2$), and a region where $\plt^{(0)} = 0$ with all
      trajectories originating on the $z=0$ axis ($\nu > 0$ and $z <
      \nu^2/2$) because of boundary condition}~{\rm
      (\ref{eq:zeroOrderBC})}. \hfill\break {\rm (b)} {\it Probability
      of escaping through the left boundary as a function of the
      incoming velocity for $\varepsilon = 0.03$ (blue line),
      $\varepsilon=10^{-2}$ (green line) and $\varepsilon = 10^{-3}$
      (red line).  We compute the trajectories using}~{\rm
      (\ref{xvdescription1})--(\ref{xvdescription2})} {\it for $\Delta
      t = 10^{-7},$ $D=1$, $L=1$, $\beta = 10^3$ and $\phibar=
      2.593$. The vertical dashed line at $- \sqrt{2 \, \phibar \,
        \beta \, D}$ separates the two cases of boundary
      condition}~{\rm (\ref{eq:kramersBC})}.}
  \label{figure1}
\end{figure}%
The solution to~\eqref{eq:zeroOrderKramersExp} is constant along the
characteristic curves
\begin{equation*}
  -\psi(z) = \frac{\nu^2}{2} + C,
\end{equation*}
for $C$ an arbitrary constant. These curves are shown in
Figure~\ref{figure1}(a) for the linear potential~(\ref{eq:linPotential}).
The curve 
\begin{equation*}
  1 -\psi(z) = \frac{\nu^2}{2}
\end{equation*}
divides the half-plane into three regions.  In the region where
$\nu > 0$ and $1-\psi(z) < \nu^2/2$, $\plt^{(0)}$ is zero due to the
boundary condition~\eqref{eq:zeroOrderBC}. \comment{This region
  corresponds to particles originating at the boundary and moving into
  the domain. Since particles are only adsorbed at the boundary, and
  not emitted, we find that the solution is zero throughout the
  region.}  Where $\nu < 0$ and $1-\psi(z) < \nu^2/2$ the solution
will be determined by matching to the outer solution. \comment{This
  region corresponds to particles entering from outside the boundary
  layer with sufficient velocity to escape over the potential barrier,
  and thereby exit through the reactive surface at $z=0$.}  Finally,
in the region $1-\psi(z) > \nu^2/2$, trajectories with $\nu < 0$ are
reflected in a symmetric manner. For points with $\nu < 0$ the
solution will again be given by matching to the outer solution, while
for those with $\nu > 0$ the solution is determined by
reflection. \comment{This region corresponds to particles that enter
  from outside the boundary layer with insufficient velocity to escape
  over the potential barrier. These particles are then reflected,
  moving back out of the boundary layer.}  Note, in the original
variables the curve separating these three regions is given by
\begin{equation*}
1-\psi\!\paren{\frac{r-\rb}{\varepsilon}} 
= \frac{(\vv \cdot \vxhat)^2}{2 \, \phibar \, \beta \, D} 
= \frac{v_r^2}{2 \, \phibar \, \beta \, D}. 
 \end{equation*}
For $\vx \in \dB_{\rb}(\vO)$, we match the inner solution as $z$
approaches the edge of the boundary layer to the outer solution as
the particle's spatial position, $\vy$, approaches $\vx$. That is, we
require
\begin{equation*}
\lim_{z \to 1} \plt^{(0)}(z,\vx,\nu,\vv,t) 
= \lim_{\vy \to \vx} p^{(0)}(\vy,\vv,t).
\end{equation*}
Using that $\psi(1)=0$, this matching condition implies the outer
solution satisfies the following reactive boundary condition 
for $v_r \geq 0$ and $\vx \in \dB_{\rb}(\vO)$
\begin{align} \label{eq:kramersBC}
p^{(0)}(\vx,\vv,t) = 
\begin{cases}
p^{(0)}(\vx,\vv - 2 \, v_r \, \vxhat, t), 
& 0 \leq v_r < \sqrt{2 \, \phibar \, \beta \, D}, \\
    0, & v_r > \sqrt{2 \, \phibar \, \beta \, D}.
  \end{cases}
\end{align}
When the moving particle reaches the reactive boundary with radial
velocity in the outward direction greater than
$\sqrt{2 \, \phibar \, \beta \, D}$ it leaves the domain (i.e.
undergoes reaction). In contrast, when the particle reaches the
boundary with a slower radial velocity in the outward direction it is
reflected back into the domain along the direction of the normal at
the point where it hit the reactive boundary. This ``specular
reflection'' boundary condition is also obtained in the corresponding
deterministic Newtonian mechanics model. We demonstrate this
explicitly for a simple one-dimensional example in
Appendix~\ref{ap:newtMechEx}.

A version of this boundary condition, given in terms of an arbitrary
threshold velocity, is assumed in the kinetic boundary layer
investigations of the one-dimensional Kramer's equation
in~\cite{Burschka:SelectAbsorpBC1981do,Burschka:FPKUnifField1982kr},
and the 3D spherically-symmetric steady-state Kramer's equation
in~\cite{Kneller:DiffControlRx1985}. It is also (briefly) mentioned
in~\cite{Kneller:DiffControlRx1985} that the specific threshold of
$\sqrt{2 \phibar \beta D}$ we derive is what one might impose across
an interface where the potential is discontinuous with a jump of size
$\phibar$ (again, for the 3D spherically-symmetric steady-state
Kramer's equation). Our asymptotic analysis shows hows this specific
threshold velocity arises in the general Kramer's equation as the
limit of a shrinking potential boundary layer bordering a Dirichlet
boundary condition. We obtain an effective jump in potential at the
reactive boundary, as opposed to an interface within the domain.
Contrast this to the limit of the BD problem from
Section~\ref{secBDgen}, where in taking $\varepsilon \to 0$ with
$\phibar$ fixed the influence of the potential is completely lost
(e.g. a zero Dirichlet boundary condition is recovered).

We therefore conclude that the LD model with the interaction potential
is in the limit that $\varepsilon \to 0$ equivalent to solving the
(zero-potential) Kramers equation
\begin{equation} 
\label{eq:kramersEqDeltaZero}
\PD{p}{t} 
+ 
\vv \cdot \nabla_{\vx} p 
= 
\beta \, \nabla_{\vv} \cdot \brac{
\vv \, p + \beta \, D \, \nabla_{\vv} p \,}, 
\qquad \quad \mbox{for} \; \vx \in \Omega, \vv \in R^d,
\end{equation}
with the specular reflection reactive boundary
condition~\eqref{eq:kramersBC} on $\dB_{\rb}(\vO)$ and whatever
boundary condition was imposed on $\partial \Omega$. \comment{That is,
  in the limit that the width of the potential approaches zero, with
  the barrier height held fixed, we find that the potential can be
  approximated by a velocity threshold boundary condition. Here
  particles moving sufficiently fast relative to the height of the
  potential barrier undergo bimolecular reactions when reaching the
  reactive boundary, while those moving too slow are reflected back
  into the domain. This result should be applicable for general
  short-range potential interactions that form a high barrier as the
  separation between two molecules approaches a fixed
  ``reaction-radius'', and then a deep well once this barrier is
  surmounted.  In contrast to the diffusive case, taking the barrier
  height $\phibar \to \infty$ (with $\beta$ fixed) leads to a complete
  loss of reaction; all particles reaching the reactive boundary are
  simply reflected back into the domain.}

\subsection{A numerical example showing recovery of boundary
condition~{\rm (\ref{eq:kramersBC})} as $\varepsilon \to 0$}
\label{secnum1}

\noindent 
We consider the LD model~(\ref{eq:langevinEq}) for $d=1$ and the
linear potential~(\ref{eq:linPotential}). In the one-dimensional case,
our computational domain is interval $\Omega=[0,L]$.  We choose a
small time step $\Delta t$ and compute the position $X(t+\Delta t)$
and the velocity $V(t+\Delta t)$ from the position $X(t)$ and the
velocity $V(t)$ by
\begin{eqnarray}
X(t+\Delta t) & = & X(t) + V(t) \, \Delta t, 
\label{xvdescription1} \\
V(t+\Delta t) & = & V(t) - \beta \, V(t) \, \Delta t 
+ 
\frac{D \, \beta \, \phibar}{\varepsilon} 
\, \chi_{[0,\varepsilon]}(X(t)) 
\, \Delta t
+
\beta \sqrt{2 D \Delta t} \; \xi,
\label{xvdescription2}
\end{eqnarray}
where $\chi_{[0,\varepsilon]}: {\mathbb R} \to \{0,1\}$
is the characteristic function of the interval $[0,\varepsilon]$
and $\xi$ is a normally distributed random variable with zero 
mean and unit variance. We implement adsorbing boundaries at both 
ends $x=0$ and $x=L$ of the simulation domain $[0,L]$ by terminating
the computed trajectory whenever $X(t) < 0$ or $X(t) > L$. 

In this paper, we are interested in understanding the dependence of 
the behavior of the LD model~(\ref{eq:langevinEq}) on its parameters 
$\varepsilon$, $\phibar$ and $\beta$. In particular, we choose the 
values of other parameters equal to 1, namely
\begin{equation}
D = L = 1.
\label{valuesofDL}
\end{equation}
In this section, we are interested in the limit $\varepsilon \to 0$.
We want to illustrate the boundary condition~(\ref{eq:kramersBC}).
Therefore we fix the values of $\phibar$ and $\beta$ and simulate
the LD model~(\ref{eq:langevinEq}) for different values of $\varepsilon$.

For each value of $\varepsilon$, we compute many trajectories 
according to~(\ref{xvdescription1})--(\ref{xvdescription2}),
starting from the middle of the domain, i.e. $X(0)=L/2$, with initial 
velocity $V(0)$ sampled from the normal distribution with zero mean 
and variance $\beta \, D.$ Whenever a particle enters the region
$[0,\varepsilon]$, we record its incoming velocity. Then we follow 
its trajectory in the region $[0,\varepsilon]$ and record one of two 
possible outputs:

{
\leftskip 5mm

\noindent
(1) the particle leaves $\Omega$ through its left boundary (i.e. $X(t)<0$); 
or 

\noindent
(2) the particle returns back to the region $(\varepsilon,L]$ of 
domain $\Omega$ (i.e. $X(t) > \varepsilon$). 

\par

\leftskip 0mm
}

\noindent
The fraction of particles which left the domain $\Omega$ through its
left boundary as a function of the incoming velocity is plotted in
Figure~\ref{figure1}(b). This curve can be interpreted
as the probability that the particle escapes over the potential
barrier, given its incoming velocity. We also plot the vertical
dashed line at $-\sqrt{2 \, \phibar \, \beta \, D}$ in 
Figure~\ref{figure1}(b). This threshold separates the two cases of 
boundary condition~(\ref{eq:kramersBC}). We observe that the 
probability of escape converges to the step function as 
$\varepsilon \to 0$, i.e. we have numerically confirmed boundary 
condition~(\ref{eq:kramersBC}).

We will return to this example in Section~\ref{secnum2} when
we study the convergence of the LD model to the diffusion process
with a Robin boundary condition. In particular, we use
values $\beta = 10^3$ and $\phibar = 2.593$ in 
Figure~\ref{figure1}(b). This choice of $\phibar$
will be explained in the following section and then used
again in one of our simulations presented 
in Section~\ref{secnum2}.

\section{From Langevin Dynamics to Brownian Dynamics}
\label{secBDLD}

\noindent
We now study the overdamped limit of the LD
model~(\ref{eq:langevinEq}).  We wish to show that in the overdamped
limit where $\beta \to \infty$, taking $\phibar \to \infty$ in a
$\beta$-dependent manner will recover a Robin boundary condition for
the limiting diffusion equation. To study the $\beta \to \infty$
limit, we extend the asymptotic analysis of the one-dimensional
Kramers equation in~\cite{Erban:2007:RBC} to the $d$-dimensional
(zero-potential) Kramers equation~(\ref{eq:kramersEqDeltaZero}). The
kinetic boundary layer studies
in~\cite{Burschka:SelectAbsorpBC1981do,Burschka:FPKUnifField1982kr,Kneller:DiffControlRx1985}
previously investigated the relationship between the velocity
threshold in the specular reflection boundary condition and effective
adsorption rate, $K$, in the Robin condition.  Our approach here
differs from those studies, which primarily used numerical solutions
of truncated moment equations or basis function expansions to estimate
empirically determined formulas for the Robin constant, $K$. We focus
on deriving an explicit formula relating how the potential barrier
height, $\phibar$, should be chosen as $\beta \to \infty$ to recover a
specified Robin constant.

We begin by re-scaling velocity as $\vv = \sqrt{\beta} \, \veta$ and
let $f(\vx,\veta,t) = p(\vx,\vv,t)$. Substituting
into~(\ref{eq:kramersEqDeltaZero}), we obtain
\begin{equation} 
\label{eq:kramersEqRescaledBeta}
\PD{f}{t} 
+ 
\sqrt{\beta} \, \veta \cdot \nabla_{\vx} f = \beta \, \nabla_{\veta} \cdot
\brac{ \veta \, p +  D\, \nabla_{\veta} p}.
\end{equation}
We expand $f$ in powers of $\beta^{-1/2}$ as
\begin{equation*}
f(\vx,\veta,t) \sim f_0(\vx,\veta,t) 
+ 
\frac{1}{\sqrt{\beta}} \, f_1(\vx,\veta,t) 
+ 
\frac{1}{\beta} \, f_2(\vx,\veta,t) + \dots.
\end{equation*}
Substituting into~\eqref{eq:kramersEqRescaledBeta}, we find
\begin{align}
\nabla_{\veta} \cdot \brac{ \veta \, f_0 + D \, \nabla_{\veta} f_0} &= 0,
\label{eq:f0Eq} \\
\nabla_{\veta} \cdot \brac{ \veta \, f_1 + D \, \nabla_{\veta} f_1} 
&= 
\veta \cdot \nabla_{\vx} f_0, \label{eq:f1Eq} \\
\nabla_{\veta} \cdot \brac{ \veta \, f_2 + D \, \nabla_{\veta} f_2} 
&= 
\veta \cdot \nabla_{\vx} f_1 + \PD{f_0}{t}. \label{eq:f2}
\end{align}
Implicit in these equations is the assumption that we are interested 
in timescales for which
\begin{equation*}
  t \gg \frac{1}{\beta}.
\end{equation*}
We therefore interpret~\eqref{eq:f0Eq} as implying that the velocity
distribution component of $f_0$ relaxes to equilibrium on a faster
timescale than $t$. Denote by $\tau$ this faster timescale. As
discussed in the introduction to~\cite{Desvillettes:2001hy}, up to a
normalization constant there is a unique solution to~\eqref{eq:f0Eq}
that corresponds to the equilibrium solution of the fast-timescale,
time-dependent equation
\begin{equation*}
  \PD{f_0}{\tau} = \nabla_{\veta} \cdot \brac{ \veta \, f_0 + D \, \nabla_{\veta} f_0}.
\end{equation*}
This equilibrium solution is then
\begin{equation*}
f_0(\vx,\veta,t) = \varrho(\vx,t) \, \exp \!\left[-\frac{\abs{\veta}^2}{2D} \right],
\end{equation*}
where $\varrho(\vx,t)$ is independent of $\veta$. We similarly find
that the general solution to~(\ref{eq:f1Eq}) is given by
\begin{equation*}
f_1(\vx,\veta,t) = \Big( -\nabla_{\vx} \varrho(\vx,t) \cdot \veta + \xi(\vx,t) \Big)
\, \exp \!\left[-\frac{\abs{\veta}^2}{2D} \right],
\end{equation*}
where $\xi(\vx,t)$ is also independent of $\veta$.  Substituting these
into~\eqref{eq:f2} we see that
\begin{equation*}
\nabla_{\veta} \cdot \brac{ \veta \, f_2 + D \, \nabla_{\veta} f_2} 
= 
\left(
\PD{\varrho}{t} 
- 
\sum_{i,j=1}^d 
\frac{\partial^2 \varrho}{\partial x_i \partial x_j} \eta_i \eta_j
+ \veta \cdot \nabla_{\vx} \xi(\vx,t) \right)
\exp \!\!\left[-\frac{\abs{\veta}^2}{2D} \right].
\end{equation*}
Integrating both sides of this equation for all $\veta \in \R^d$, and
using that $f_2(\vx,\veta,t) \to 0$ and 
$\abs{\nabla_{\veta} f_2(\vx,\veta,t)} \to 0$ 
as $\abs{\veta} \to \infty$, we conclude
that $\varrho(\vx,t)$ satisfies the diffusion equation
\begin{equation*}
  \PD{\varrho}{t} = D \lap_{\vx} \varrho.
\end{equation*}
The probability density that the particle has position $\vx$ at time
$t$ and has not reacted is given by
\begin{equation*}
  u(\vx,t) = \int_{\R^d} p(\vx,\vv,t) \, \mbox{d}\vv,
\end{equation*}
implying that to leading order $u$ is proportional to $\varrho$. As
such, we expect as $\beta \to \infty$ that $u$ satisfies the diffusion
equation.

We now show that $\varrho$, and hence $u$, satisfies a Robin boundary
condition as $\beta \to \infty$ when $\phibar$ is chosen to approach
infinity in a $\beta$-dependent manner. To leading order in $\beta$,
the outward flux through a point, $\vx \in \partial B_{\rb}(\vO)$, is
\begin{align}
\nonumber
J(\vx,t) 
:&= 
-\int_{\R^d} (\vv \cdot \vxhat) \, p(\vx,\vv,t) \, \mbox{d} \vv \\
\nonumber
&\sim -\int_{\R^d} (\vv \cdot \vxhat) \brac{f_0(\vx,\vv/\sqrt{\beta},t) 
+ \frac{1}{\sqrt{\beta}} \, f_1(\vx,\vv/\sqrt{\beta},t)} 
\, \mbox{d} \vv\, \\
\label{genflux}
&= (2 \, \pi \, D \, \beta)^{d/2} 
\Big( 
D \, \nabla_{\vx} \varrho(\vx,t) \cdot \vxhat
\Big),
\end{align}
where as in previous sections $\vxhat = \vx / \abs{\vx}$. Let
$$
\mathcal{R}(\vxhat) 
= 
\left\{ \vv \, \big| 
\, \vv \cdot \vxhat < - \sqrt{2 \, \phibar \, \beta \, D } \right\}
= 
\left\{ \vv \, \big| 
\, v_r < - \sqrt{2 \, \phibar \, \beta \, D } \right\}
$$ 
denote the set of velocities at which particles may escape through the
reactive boundary. The specular reflection boundary
condition~(\ref{eq:kramersBC}) implies
\begin{align*}
J(\vx,t) &= 
-\int_{\mathcal{R}(\vxhat)} \paren{\vv \cdot \vxhat} p(\vx,\vv,t) 
\, \mbox{d}\vv \\
&\sim -\int_{\mathcal{R}(\vxhat)}\paren{\vv \cdot \vxhat} 
\brac{ f_0(\vx,\vv/\sqrt{\beta},t) 
+ 
\frac{1}{\sqrt{\beta}} \, f_1(\vx,\vv/\sqrt{\beta},t)} \, \mbox{d}\vv 
\\
&= (2 \, \pi \, D \, \beta)^{d/2}
\left[ \sqrt{\frac{\beta \, D}{2 \pi}} \, \exp [-\phibar] \, \varrho(\vx,t) 
+ \sqrt{\frac{D}{2 \pi}} \, \exp [-\phibar] \, \xi(\vx,t) \right. \\
& \qquad \qquad \qquad \qquad \qquad + 
\left. \paren{
\sqrt{
\frac{\phibar}{\pi}} 
\, \exp [-\phibar] 
+ 
\frac{\erfc(\sqrt{\phibar})}{2}
} 
\Big( 
D \, \nabla_{\vx} \varrho(\vx,t) \cdot \vxhat
\Big)  \right],
\end{align*}
for $\vx \in \dB_{\rb}(\vO)$ and $\vxhat = \vx/\rb$.
Comparing with~(\ref{genflux}), we find that to leading order
  $\varrho$, and hence $u$, satisfy the reactive Robin boundary
  condition for $\beta \to \infty$
\begin{equation*}
D \, \nabla_{\vx} \varrho(\vx,t) \cdot \vxhat
= 
\sqrt{\beta} \, \overline{K}(\phibar,D) \, \varrho(\vx,t)
\qquad \quad \mbox{for} \; \vx \in \dB_{\rb}(\vO),
\end{equation*}
where
\begin{equation}
\overline{K}(\phibar,D)
= 
\frac{\displaystyle \sqrt{\frac{D}{2 \, \pi}} 
\, \exp [-\phibar]}
{\displaystyle
1 
- 
\sqrt{
\frac{\phibar}{\pi}} 
\, \exp [-\phibar] 
- 
\frac{\erfc(\sqrt{\phibar})}{2} 
}.
\label{KbetaphiDequation}
\end{equation}
As $\phibar \to \infty$, the denominator of $\overline{K}$ approaches
one. This suggests the scaling
\begin{equation}
\phibar 
= 
\ln \paren{ 
\frac{1}{K}
\sqrt{\frac{\beta \, D}{2 \, \pi}}
},
\label{phibarscaling}
\end{equation}
so that 
\begin{align*}
\sqrt{\beta} \, \overline{K}(\phibar,D) &\to K,
\qquad \qquad \mbox{as} \; \beta \to \infty.
\end{align*}
We then find $\varrho(\vx,t)$ satisfies the desired Robin boundary
condition,
\begin{equation*}
D \, \nabla_{\vx} \varrho(\vx,t) \cdot \vxhat =  K \, \varrho(\vx,t)
\qquad \quad \mbox{for} \; \vx \in \dB_{\rb}(\vO),
\end{equation*}
in the limit $\beta \to \infty$.  More generally, we could impose the
equality $\sqrt{\beta} \, \overline{K}(\phibar,D)=K$ in
equation~(\ref{KbetaphiDequation}). Solving for $\beta$, we obtain the
following relation between $\phibar$ and $\beta$:
\begin{equation}
\beta
=
\frac{K^2}{D} \left( 
\sqrt{2 \, \pi}
\, \exp [\phibar]
- 
\sqrt{2 \, \phibar} 
- 
\sqrt{\frac{\pi}{2}}
\,
\erfc(\sqrt{\phibar}) \, \exp [\phibar]
\right)^2.
\label{betaphibarrelation}
\end{equation}
In Figure~\ref{figure2}(a), we compare both
formulae~(\ref{phibarscaling}) and (\ref{betaphibarrelation}) for
$K=D=1$. As expected, they are equivalent in the limit
$\beta \to \infty.$ In the following section, we will present
illustrative simulations, confirming that~(\ref{betaphibarrelation})
provides a more accurate approximation of the limiting Robin boundary
condition. Further improvements to these formulas could presumably be
made by incorporating a more detailed representation of the kinetic
boundary layer near $\partial B_{\rb}(\vO)$.

\subsection{A numerical example illustrating limit $\beta \to \infty$}
\label{secnum2}

\noindent 

As in Section~\ref{secnum1}, we consider the LD model~(\ref{eq:langevinEq})
for $d=1$  in the interval
$\Omega=[0,L]$ with the linear potential~(\ref{eq:linPotential}) near the
left-hand boundary only. 
We choose a small time step $\Delta t$ and compute the 
position $X(t+\Delta t)$ and the velocity $V(t+\Delta t)$ from 
the position $X(t)$ and the velocity $V(t)$ by
(\ref{xvdescription1})--(\ref{xvdescription2}). We implement adsorbing
boundaries at both ends $x=0$ and $x=L$ of the simulation domain $[0,L]$ 
by terminating the computed trajectory whenever $X(t) < 0$ or $X(t) > L$. 

\begin{figure}[t]
\relax
\vskip 2mm
\centerline{
\hskip 2mm
\includegraphics[width=0.48\columnwidth]{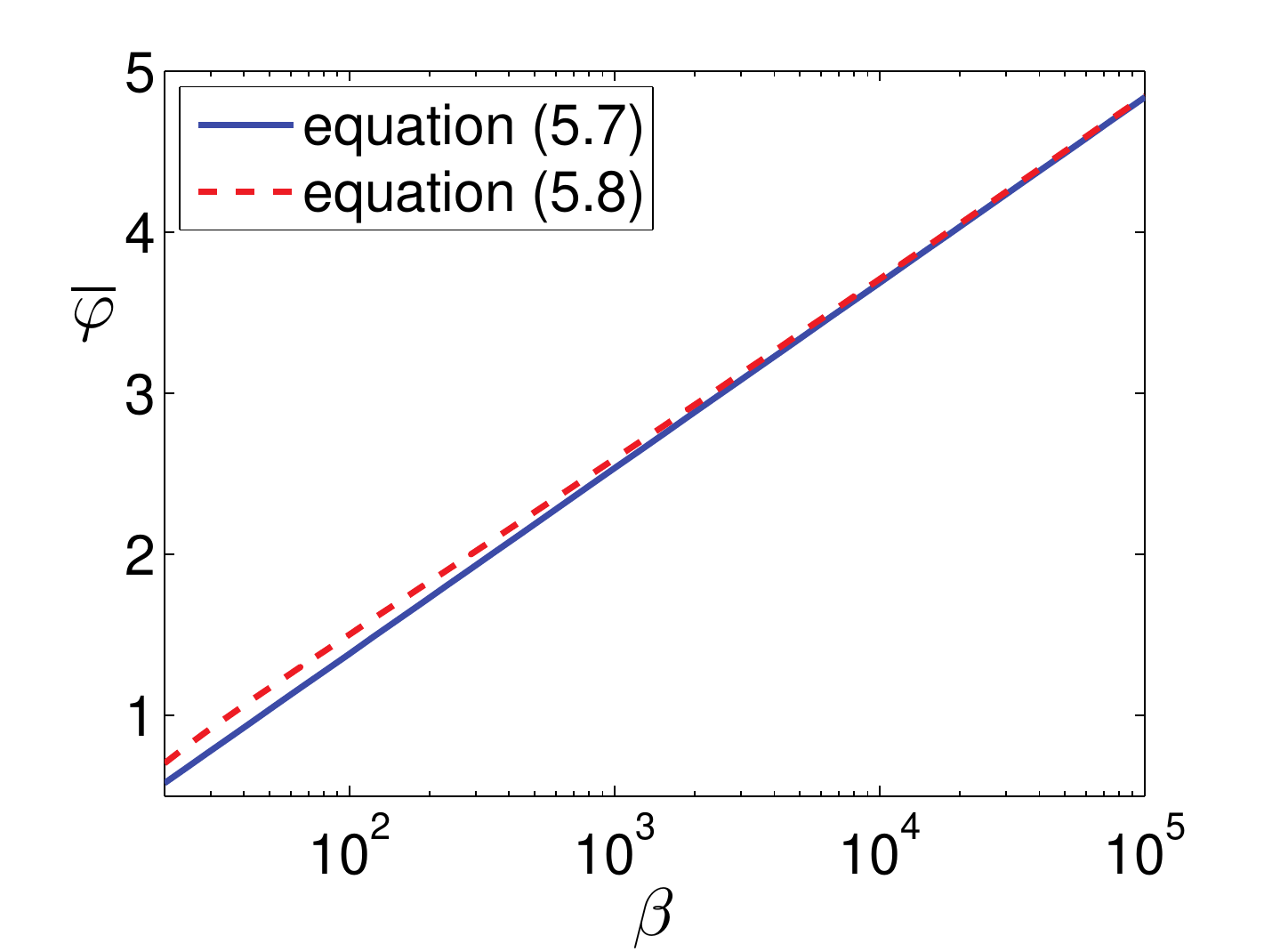}
\hskip 4mm
\includegraphics[width=0.48\columnwidth]{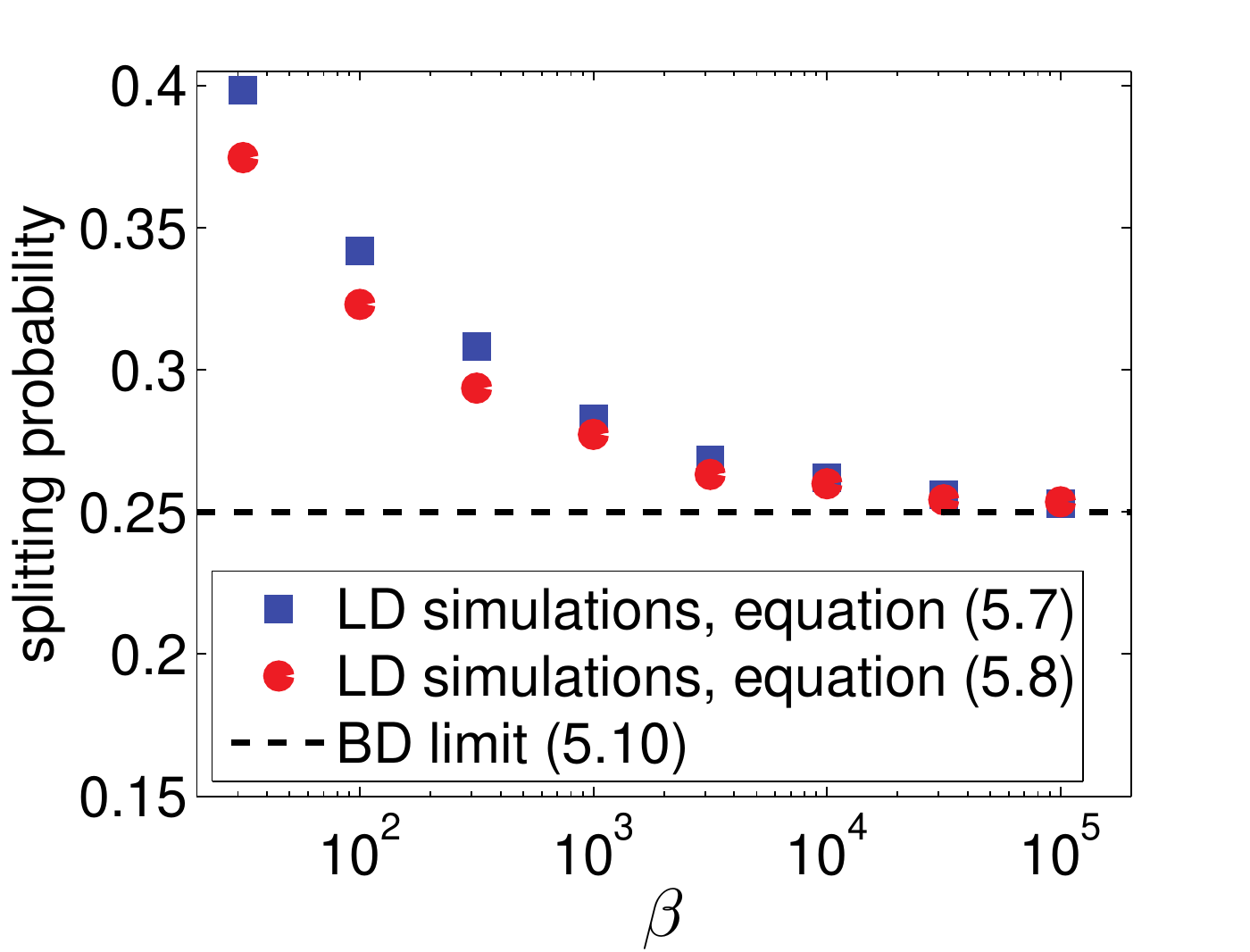}
}
\vskip -4.9cm
\leftline{\hskip 1mm (a) \hskip 6.1cm (b)}
\vskip 4.3cm
\caption{{\rm (a)} {\it Dependence of $\phibar$ on $\beta$ computed by
    equation}~{\rm (\ref{phibarscaling})} {\it (blue solid line)
    and equation}~{\rm (\ref{betaphibarrelation})} {\it (red dashed line)
    for $K=D=1$.} \hfill\break {\rm (b)} {\it Splitting probability of
    escaping through the left boundary as a function $\beta$. In each
    simulation, we estimate the splitting probability as an
    average over $10^5$ realizations. We use
    $\varepsilon = 10^{-3}$, $D=L=K=1$ and
    $\phibar$ computed according to equation}~{\rm
    (\ref{phibarscaling})} {\it (blue squares) or equation}~{\rm
    (\ref{betaphibarrelation})} {\it (red circles).  The dashed line
    denotes the theoretical BD result}~{\rm
    (\ref{splitprobtheory})}.}
\label{figure2}
\end{figure}%
In this section, we want to show that both 
equations~(\ref{phibarscaling}) and~(\ref{betaphibarrelation}) 
correctly recover the Robin boundary condition in the limit
$\beta \to \infty.$  In particular, we choose the 
values of other parameters equal to 1, namely (compare 
with~(\ref{valuesofDL})) 
\begin{equation}
K = D = L = 1.
\label{valuesofDLK}
\end{equation}
We vary $\beta$ and we use either equation~(\ref{phibarscaling}) 
or equation~(\ref{betaphibarrelation}) to calculate the corresponding 
value of $\phibar$ (these values are plotted in Figure~\ref{figure2}(a)).

For each value of $\beta$, we compute $10^5$ trajectories 
according to (\ref{xvdescription1})--(\ref{xvdescription2}),
starting from the middle of the domain, i.e. $X(0)=L/2$, with initial 
velocity $V(0)$ sampled from the normal distribution with zero mean 
and variance $\beta \, D.$ Each trajectory is calculated until
it leaves the domain $[0,L]$ either through the left or right
boundary point. 
 In Figure~\ref{figure2}(b), we plot
the  probability that a trajectory leaves the 
domain $\Omega$ through the left boundary (the so-called splitting
probability), estimated 
as the fraction of all trajectories which are terminated because 
$X(t) < 0$.

Our goal is to illustrate that equations~(\ref{phibarscaling}) 
and (\ref{betaphibarrelation}) can be used to connect the LD model 
with the limiting Robin boundary
problem~(\ref{diffplusrobin1})--(\ref{diffplusrobin2}).
Let $\Pi(x)$ be the  probability that the BD
particle leaves the domain $\Omega = [0,L]$ through the left 
boundary. Since
$$
\frac{\mbox{d}^2 \Pi}{\mbox{d} x^2}(x) = 0, \quad
\mbox{for} \; x \in \Omega, 
\qquad
\mbox{with}
\qquad
D \frac{\mbox{d} \Pi}{\mbox{d} x}(0) = K \, (\Pi(0) - 1)
\qquad
\mbox{and}
\qquad
\Pi(L)=0,
$$
we find
$$
\Pi(x) = \frac{K \, (L-x)}{D + KL}.
$$
Since all trajectories  start from the middle of the domain,
$X(0)=L/2$, we have
\begin{equation}
\Pi\!\left(\frac{L}{2}\right) 
= 
\frac{K \, L}{2 (D + KL)}
=
0.25 \%,
\label{splitprobtheory}
\end{equation}
for the parameter values given by (\ref{valuesofDLK}).
This value is plotted in Figure~\ref{figure2}(b) as the
black dashed line. We confirm that the results estimated from
simulations approach~(\ref{splitprobtheory}) as $\beta \to \infty$.
We also confirm that simulations based on the higher-order 
approximation~(\ref{betaphibarrelation})
 converge more quickly to the limiting Robin boundary problem than
the simulations based on equation~(\ref{phibarscaling}).

\section{Discussion}
\label{secdiscussion}

\noindent
We have considered three parameters, $\varepsilon$ (potential width),
$\phibar$ (potential height) and $\beta$ (friction constant),
and studied several limits of these parameters which lead to the
Robin (reactive) boundary condition~(\ref{diffplusrobin2}).
Parameters $\varepsilon$ and $\phibar$ are shared by both the BD
and LD models. In Section~\ref{secBD}, we have shown that
the BD model can recover the Robin boundary condition 
in the limit $\varepsilon \to 0$ and $\phibar \to \infty$ 
when these parameters are related by (\ref{epsscalinggenBD}). For the case of  
the  linear potential~(\ref{eq:linPotential}) this
relation can be rewritten as
\begin{equation}
\phibar 
-
\ln(\phibar)
= 
\ln \paren{
\frac{D}{K \varepsilon}
}.
\label{BDresult}
\end{equation}
The LD model has an additional parameter $\beta$. 
In Section~\ref{secBDLD}, we have derived two formulae
(\ref{phibarscaling}) and (\ref{betaphibarrelation})
which relate the LD model to the BD model with the Robin boundary 
condition~(\ref{diffplusrobin2}). Both results, 
(\ref{phibarscaling}) and (\ref{betaphibarrelation}),
are equivalent in the limit $\beta \to \infty$. 
Equation~(\ref{betaphibarrelation}) is more accurate for
finite values of $\beta$, while~(\ref{phibarscaling})
is simpler and easier to interpret. It is given as
\begin{equation}
\phibar 
= 
\ln \paren{ 
\frac{1}{K}
\sqrt{\frac{\beta \, D}{2 \, \pi}}
}.
\label{LDresult}
\end{equation}
Considering that (experimentally determinable) parameters $K$ and $D$
are given constants, we can compare our BD result~(\ref{BDresult})
with our LD result~(\ref{LDresult}).  They can be both used to specify
the height of the potential barrier $\phibar$, which is given as a
function of $\varepsilon$ in~(\ref{BDresult}) and as a function of
$\beta$ in~(\ref{LDresult}).

On the face of it, the LD result~(\ref{LDresult}) does not depend on
the parameter $\varepsilon$. However, the derivation of the LD
result~(\ref{LDresult}) is only valid for small $\varepsilon$. More
precisely, our conclusion for the LD model~(\ref{eq:langevinEq}) can
be stated as follows:

\smallskip

{\leftskip 5mm

\noindent
(1) If $\varepsilon \ll \sqrt{\frac{D}{\beta}} \ll 1$, then
$\phibar$ is independent of $\varepsilon$ and can be written in 
the form~(\ref{LDresult});

\noindent
(2) If $1 \gg \varepsilon \gg \sqrt{\frac{D}{\beta}}$, then $\phibar$ is
independent of $\beta$ and is given by~(\ref{BDresult}).

\par

\leftskip 0mm
}

\smallskip

\noindent
In our illustrative computations in Figure~\ref{figure2}(b) we have
used $\varepsilon = 10^{-3}$. In this case, the BD
result~(\ref{BDresult}) implies that $\phibar =9.118$. We observe that
this value is higher than the values of $\phibar$ plotted in
Figure~\ref{figure2}(a) which are used for our simulations in
Figure~\ref{figure2}(b).  We can also substitute this value
$\phibar =9.118$ into~(\ref{LDresult}).  We obtain
$\beta = 5.224 \times 10^8$. For these values of $\beta$ and
$\varepsilon$ we are in case~(2), so that the LD
result~(\ref{LDresult}) would no longer be applicable. We instead
recover the limiting Robin boundary condition (in the limit
$\beta \to \infty$ with $\varepsilon = 10^{-3}$) by using the BD
result~(\ref{BDresult}).


The above results~(\ref{BDresult})--(\ref{LDresult}) can be used to
connect the experimentally determinable parameters $K$ and $D$ with
parameters of computer simulations to design reaction-diffusion models
based on LD, i.e. to simulate diffusion-limited bimolecular reactions
between Kramers particles. There are also other situations where our
analysis will be applicable. One of them is adsorption to
surfaces~\cite{Erban:2007:DPI,Erban:2007:TSR}. Our set up includes
interactions with a reactive surface of a sphere and can be used in
modeling interactions of small molecules with large reactive spheres,
for example, for adsorption of polymers to a surface of a
virus~\cite{Erban:2007:DPI} or for coating of spherical particles by
reactive polymers~\cite{Subr:2006:CDC}. We also expect our results
should be easy to extend to general non-spherical reaction surfaces,
assuming sufficient regularity.  Another possible application area is
modeling excluded-volume effects. We have observed that the short
range repulsive interaction potential~(\ref{phidef}) leads to zero
Neumann boundary condition~(\ref{neumannboundary}) if $\varepsilon$ is
chosen to approach zero slower than~(\ref{epsscalinggenBD}) (in the
limit $\phibar \to \infty$). A similar potential mechanism has been
used to enforce Neumann boundary conditions on global domain boundaries
in~\cite{AtzbergerOsmosis2009}, and can be used to model
excluded-volume effects in models of intracellular macro-molecular
crowding~\cite{Arjunan:2010:NMR,Bruna:2012:EED}.

\comment{Since LD requires a smaller time step than the overdamped 
BD model, the numerical simulations are in general more computationally intensive for LD.} However, for many biological applications the
LD model is only required close to a reactive surface (where we have a
non-zero potential). In particular, one could replace the
computationally intensive LD model with the BD model in the part of
the computational domain which is far from the reactive
surface~\cite{Erban:2014:MDB}. In some applications, one could further
substitute the BD simulation algorithms by even coarser and more
efficient simulation techniques, including lattice-based
simulations~\cite{Flegg:2012:TRM,Engblom:2009:SSR} or even mean-field
equations~\cite{Erban:2007:TSR,Franz:2013:MRA}.  In this way, one
could design LD simulation methods which simulate intracellular
processes on comparable time scales as the BD simulation packages
which are available in the
literature~\cite{Andrews:2004:SSC,Robinson:2015:MRS,Takahashi:2010:STC}.
In some applications, one could also design an efficient
first-passage-time scheme by replacing
discretization~(\ref{xvdescription1})--(\ref{xvdescription2}) (which
uses fixed time step $\Delta t$) by estimating the next time when a
Kramers particle becomes sufficiently close to the reactive
target~\cite{Hagan:1989:MET}. Similar approaches have been used to
accelerate BD simulations in the
literature~~\cite{vanZon:2005:GFR,Opplestrup:2009:FKM,Mauro:2014fi}.

\appendix
\section{A simple example illustrating how specular reflection arises in a classical mechanical system}
\label{ap:newtMechEx}
In this section we illustrate with a simple example how the specular
reflection boundary condition~\eqref{eq:kramersBC} arises in a
classical mechanical system in the absence of noise.

We consider the simple case of a particle moving in the infinite
one-dimensional potential,
\begin{equation*}
  \kb T \varphi(x) =
  \begin{cases}
    -\frac{\kb T \phibar}{\varepsilon} x, & x \leq 0, \\
    0, & x > 0,
  \end{cases}
\end{equation*}
and experiencing friction, with friction constant $\beta$. Note, for
the purposes of this section we only make use of the value of the
potential for $x \in \left(-\infty,0\right]$.  Newton's equation when
the particle's position, $X(t)$, satisfies $X(t) \leq 0$ with negative
velocity is then
\begin{equation*}
  m\D{V}{t} = -m \beta V + \frac{\kb T \phibar}{\varepsilon}.
\end{equation*}
Using the Einstein Relation this reduces to
\begin{equation*}
  \D{V}{t} = -\beta V + \frac{D \beta \phibar}{\varepsilon},
\end{equation*}
the one-dimensional analogue of the LD model~\eqref{eq:langevinEq}
with the noise term neglected.

We consider the initial conditions
\begin{align*}
  V(0) = v_0 < 0, && X(0) = 0,
\end{align*}
and ask for what initial velocity range the molecule successfully
moves a distance $\varepsilon$ to the left before changing direction and
falling down the potential gradient. This is consistent with a
molecule reaching the reactive boundary in the Kramer's equation
model~\eqref{eq:kramersEq}. Given the molecule's negative initial
velocity, until the molecule's direction of motion reverses and the
molecule moves back across the origin we find
\begin{equation*}
  V(t) = \paren{v_0 - \frac{D\phibar}{\varepsilon}} \exp\brac{-\beta t} + \frac{D\phibar}{\varepsilon}.
\end{equation*}
The velocity then becomes zero at time
\begin{equation*}
  t^* = \frac{1}{\beta} \ln \paren{1 - \alpha},
\end{equation*}
where 
\begin{equation*}
  \alpha = \frac{\varepsilon v_0}{D \phibar} < 0. 
\end{equation*}
At this time, the molecule's position is
\begin{align*}
  X(t^*) &= \frac{1}{\beta} \paren{v_0 - \frac{D\phibar}{\varepsilon}} \paren{1 - \exp\brac{-\beta t^*}} + \frac{D\phibar t^*}{\varepsilon}, \\
  &= \frac{D \phibar}{\varepsilon \beta} \paren{ \alpha - \ln(1-\alpha)}.
\end{align*}
The molecule then successfully travels further than $-\varepsilon$,
analogous to penetrating the ``reactive boundary'' at $x=-\varepsilon$,
if
\begin{equation*}
  X(t^*) < -\varepsilon,
\end{equation*}
or equivalently that
\begin{equation*}
  \frac{D \phibar}{\varepsilon \beta} \paren{\alpha - \ln(1-\alpha)} < -\varepsilon.
\end{equation*}
As $\varepsilon \to 0$ 
\begin{equation*}
  X(t^*) \sim \frac{-\varepsilon v_0^2}{2 \phibar \beta D}.
\end{equation*}
We therefore find that the molecule successfully penetrates the
reactive boundary in the limit that $\varepsilon \to 0$ if and only if
\begin{equation*}
  v_0 < -\sqrt{2 \phibar \beta D},
\end{equation*}
consistent with the specular reflection boundary
condition~\eqref{eq:kramersBC}.


\begin{thebibliography}{10}

\bibitem{Agbanusi:2014:CBR}
{\sc I.~Agbanusi and S.~Isaacson}, {\em A comparison of bimolecular reaction
  models for stochastic reaction-diffusion systems}, Bulletin of Mathematical
  Biology, 76 (2014), pp.~922--946.

\bibitem{Andrews:2010:DSC}
{\sc S.~Andrews, N.~Addy, R.~Brent, and A.~Arkin}, {\em Detailed simulations of
  cell biology with smoldyn 2.1}, PLOS Computational Biology, 6 (2010),
  p.~e1000705.

\bibitem{Andrews:2004:SSC}
{\sc S.~Andrews and D.~Bray}, {\em Stochastic simulation of chemical reactions
  with spatial resolution and single molecule detail}, Physical Biology, 1
  (2004), pp.~137--151.

\bibitem{Arjunan:2010:NMR}
{\sc S.~Arjunan and M.~Tomita}, {\em A new multicompartmental
  reaction-diffusion modeling method links transient membrane attachment of
  {E}. coli {M}in{E} to {E}-ring formation}, Systems and Synthetic Biology, 4
  (2010), pp.~35--53.

\bibitem{AtzbergerOsmosis2009}
{\sc P.~J. Atzberger, S.~A. Isaacson, and C.~S. Peskin}, {\em A microfluidic
  pumping mechanism driven by non-equilibrium osmotic effects}, Physica D, 238
  (2009), pp.~1168--1179.

\bibitem{Bruna:2012:EED}
{\sc M.~Bruna and S.~Chapman}, {\em Excluded-volume effects in the diffusion of
  hard spheres}, Physical Review E, 85 (2012), p.~011103.

\bibitem{Burschka:SelectAbsorpBC1981do}
{\sc M.~A. Burschka and U.~M. Titulaer}, {\em {The kinetic boundary layer for
  the Fokker-Planck equation: Selectively absorbing boundaries}}, Journal of
  Statistical Physics, 26 (1981), pp.~59--71.

\bibitem{Burschka:FPKUnifField1982kr}
{\sc M.~A. Burschka and U.~M. Titulaer}, {\em {The kinetic
boundary layer for the Fokker-Planck equation: A Brownian particle in a
uniform field}}, Physica A, 112 (1982), pp.~315--330.

\bibitem{Desvillettes:2001hy}
{\sc L.~Desvillettes and C.~Villani}, {\em {On the trend to global equilibrium
  in spatially inhomogeneous entropy-dissipating systems: the linear
  Fokker-Planck equation}}, Communications on Pure and Applied Mathematics, 54
  (2001), pp.~1--42.

\bibitem{Dobramysl:2015:PMM}
{\sc U.~Dobramysl, S.~R\"udiger, and R.~Erban}, {\em Particle-based multiscale
  modeling of intracellular calcium dynamics}.
\newblock submitted to Multiscale Modelling and Simulation, available as
  http://arxiv.org/abs/1504.00146, 2015.

\bibitem{Engblom:2009:SSR}
{\sc S.~Engblom, L.~Ferm, A.~Hellander, and P.~L\"otstedt}, {\em Simulation of
  stochastic reaction-diffusion processes on unstructured meshes}, SIAM Journal
  on Scientific Computing, 31 (2009), pp.~1774--1797.

\bibitem{Erban:2014:MDB}
{\sc R.~Erban}, {\em From molecular dynamics to {B}rownian dynamics},
  Proceedings of the Royal Society A, 470 (2014), p.~20140036.

\bibitem{BDionref}
{\sc R.~Erban}, {\em Coupling all-atom molecular dynamics simulations 
of ions in water with {B}rownian dynamics}, submitted, preprint
available as: {http://arxiv.org/abs/1508.02805} (2015)

\bibitem{Erban:2007:RBC}
{\sc R.~Erban and S.~J. Chapman}, {\em Reactive boundary conditions for
  stochastic simulations of reaction-diffusion processes}, Physical Biology, 4
  (2007), pp.~16--28.

\bibitem{Erban:2007:TSR}
{\sc R.~Erban and S.~J. Chapman}, {\em Time scale of random
  sequential adsorption}, Physical Review E, 75 (2007), p.~041116.

\bibitem{Erban:2009:SMR}
{\sc R.~Erban and S.~J. Chapman}, {\em Stochastic modelling
  of reaction-diffusion processes: algorithms for bimolecular reactions},
  Physical Biology, 6 (2009), p.~046001.

\bibitem{Erban:2007:DPI}
{\sc R.~Erban, S.~J. Chapman, K.~Fisher, I.~Kevrekidis, and L.~Seymour}, {\em
  Dynamics of polydisperse irreversible adsorption: a pharmacological example},
  Mathematical Models and Methods in Applied Sciences (M3AS), 17 (2007),
  pp.~759--781.

\bibitem{Erban:2014:MSR}
{\sc R.~Erban, M.~Flegg, and G.~Papoian}, {\em Multiscale stochastic
  reaction-diffusion modelling: application to actin dynamics in filopodia},
  Bulletin of Mathematical Biology, 76 (2014), pp.~799--818.

\bibitem{Flegg:2012:TRM}
{\sc M.~Flegg, J.~Chapman, and R.~Erban}, {\em The two-regime method for
  optimizing stochastic reaction-diffusion simulations}, Journal of the Royal
  Society Interface, 9 (2012), pp.~859--868.

\bibitem{Franz:2013:MRA}
{\sc B.~Franz, M.~Flegg, J.~Chapman, and R.~Erban}, {\em Multiscale
  reaction-diffusion algorithms: {PDE}-assisted {B}rownian dynamics}, SIAM
  Journal on Applied Mathematics, 73 (2013), pp.~1224--1247.

\bibitem{Hagan:1989:MET}
{\sc P.~Hagan, C.~Doering, and C.~Levermore}, {\em Mean exit times for
  particles driven by weakly colored noise}, SIAM Journal on Applied
  Mathematics, 49 (1989), pp.~1480--1513.

\bibitem{KeizerJPhysChem82}
{\sc J.~Keizer}, {\em Nonequilibrium statistical thermodynamics and the effect
  of diffusion on chemical reaction rates}, J. Phys. Chem., 86 (1982),
  pp.~5052--5067.

\bibitem{Kneller:DiffControlRx1985}
{\sc G.~R. Kneller and U.~M. Titulaer}, {\em {Boundary layer effects on the
  rate of diffusion controlled reactions}}, Physica A, 129A (1985),
  pp.~514--534.

\bibitem{Leimkuhler:2015:MDD}
{\sc B.~Leimkuhler and C.~Matthews}, {\em Molecular Dynamics: with
  deterministic and stochastic numerical methods}, Springer, 2015.

\bibitem{li:2009vs}
{\sc X.~Li, J.~Lowengrub, A.~Raetz, and A.~Voigt}, {\em {Solving {PDEs} in
  Complex Geometries: a Diffuse Domain Approach}}, Comm.Math. Sci., 7 (2009),
  pp.~81--107.

\bibitem{Lipkow:2005:SDP}
{\sc K.~Lipkow, S.~Andrews, and D.~Bray}, {\em Simulated diffusion of
  phosphorylated {C}he{Y} through the cytoplasm of {E}scherichia coli}, Journal
  of Bacteriology, 187 (2005), pp.~45--53.

\bibitem{Mauro:2014fi}
{\sc A.~J. Mauro, J.~K. Sigurdsson, J.~Shrake, P.~J. Atzberger, and S.~A.
  Isaacson}, {\em A first-passage kinetic {M}onte {C}arlo method for
  reaction-drift-diffusion processes}, Journal of Computational Physics, 259
  (2014), pp.~536--567.

\bibitem{Opplestrup:2009:FKM}
{\sc T.~Opplestrup, V.~Bulatov, A.~Donev, M.~Kalos, G.~Gilmer, and B.~Sadigh},
  {\em First-passage kinetic {M}onte {C}arlo method}, Physical Review E, 80
  (2009), p.~066701.

\bibitem{pego:1989wf}
{\sc R.~L. Pego}, {\em {Front migration in the nonlinear {C}ahn-{H}illiard
  equation}}, Proc. R. Soc. Lond. Ser. A Math. Phys. Eng. Sci., 422 (1989),
  pp.~261--278.

\bibitem{Robinson:2015:MRS}
{\sc M.~Robinson, S.~Andrews, and R.~Erban}, {\em Multiscale reaction-diffusion
  simulations with {S}moldyn}, Bioinformatics,  (2015), p.~doi:
  10.1093/bioinformatics/btv149.

\bibitem{SzaboShoup1982}
{\sc D.~Shoup and A.~Szabo}, {\em Role of diffusion in ligand binding to
  macromolecules and cell-bound receptors}, Biophys. J., 40 (1982), pp.~33--39.

\bibitem{Smoluchowski:1917:VMT}
{\sc M.~Smoluchowski}, {\em Versuch einer mathematischen {T}heorie der
  {K}oagulationskinetik kolloider {L}\"osungen}, Zeitschrift f\"ur
  physikalische Chemie, 92 (1917), pp.~129--168.

\bibitem{Subr:2006:CDC}
{\sc V.~\v{S}ubr, {\v{C}}.~Ko\v{n}\'{a}k, R.~Laga, and K.~Ulbrich}, {\em
  Coating of {DNA}/{P}oly({L}-lysine) complexes by covalent attachment of
  poly[{N}-(2-hydroxypropyl)methacrylamide]}, Biomacromolecules, 7 (2006),
  pp.~122--130.

\bibitem{Takahashi:2010:STC}
{\sc K.~Takahashi, S.~Tanase-Nicola, and P.~ten Wolde}, {\em Spatio-temporal
  correlations can drastically change the response of a mapk pathway}, PNAS,
  107 (2010), pp.~19820--19825.

\bibitem{vanZon:2005:GFR}
{\sc J.~van Zon and P.~ten Wolde}, {\em Green's-function reaction dynamics: a
  particle-based approach for simulating biochemical networks in time and
  space}, Journal of Chemical Physics, 123 (2005), p.~234910.

\end{thebibliography}
\end{document}